\documentclass[12pt]{article}

\pdfoutput=1

\usepackage{cite}

\usepackage{amssymb,latexsym}

\usepackage{epstopdf}

\usepackage{graphics,epsfig}

\usepackage{color}

\usepackage{amsmath,amsfonts}

\usepackage{mathrsfs}

\DeclareMathAlphabet{\mathpzc}{OT1}{pzc}{m}{it}

\usepackage{tikz}
\usetikzlibrary{arrows,shapes}
\usetikzlibrary{trees}
\usetikzlibrary{matrix,arrows} 				
\usetikzlibrary{positioning}				
\usetikzlibrary{calc,through}				
\usetikzlibrary{decorations.pathreplacing}  
\usepackage{pgffor}							

\usetikzlibrary{decorations.pathmorphing}	
\usetikzlibrary{decorations.markings}
\tikzset{
    vector/.style={decorate, decoration={snake}, draw},
	provector/.style={decorate, decoration={snake,amplitude=2.5pt}, draw},
	antivector/.style={decorate, decoration={snake,amplitude=-2.5pt}, draw},
    fermion/.style={draw=black, postaction={decorate},
        decoration={markings,mark=at position .55 with {\arrow[draw=black]{>}}}},
    fermionbar/.style={draw=black, postaction={decorate},
        decoration={markings,mark=at position .55 with {\arrow[draw=black]{<}}}},
    fermionnoarrow/.style={draw=black},
    gluon/.style={decorate, draw=black,
        decoration={coil,amplitude=4pt, segment length=5pt}},
    scalar/.style={dashed,draw=black, postaction={decorate},
        decoration={markings,mark=at position .55 with {\arrow[draw=black]{>}}}},
    scalarbar/.style={dashed,draw=black, postaction={decorate},
        decoration={markings,mark=at position .55 with {\arrow[draw=black]{<}}}},
    scalarnoarrow/.style={dashed,draw=black},
    electron/.style={draw=black, postaction={decorate},
        decoration={markings,mark=at position .55 with {\arrow[draw=black]{>}}}},
	bigvector/.style={decorate, decoration={snake,amplitude=4pt}, draw},
}

\tikzstyle{block} = [draw, rectangle, 
    minimum height=3em, minimum width=6em]

\let\a=\alpha \let\b=\beta \let\g=\gamma \let\d=\delta \let\e=\epsilon
\let\z=\zeta  \let\th=\theta  \let\k=\kappa
\let\l=\lambda \let\m=\mu \let\n=\nu \let\x=\xi \let\p=\pi 
\let\s=\sigma   \let\f=\phi  
 
\let\w=\omega      \let\G=\Gamma \let\D=\Delta \let\Th=\Theta \let\L=\Lambda
\let\X=\Xi  \let\S=\Sigma  \let\Y=\Psi
 
\let\la=\label  
  
\def\nn{\nonumber} \def\bd{\begin{document}} \def\ed{\end{document}}
\def\ds{\documentstyle} \let\fr=\frac \let\bl=\bigl \let\br=\bigr
\let\Br=\Bigr \let\Bl=\Bigl
\let\bm=\bibitem
\let\na=\nabla
\def\tU{{\widetilde U}}
\let\pa=\partial \let\ov=\overline
\def\ie{{\it i.e.\ }}
\newcommand{\be}{\begin{equation}}
\newcommand{\ee}{\end{equation}}
\def\ba{\begin{array}}
\def\ea{\end{array}}
\def\ft#1#2{{\textstyle{{\scriptstyle #1}\over {\scriptstyle #2}}}}
\def\fft#1#2{{#1 \over #2}}
\def\F#1#2{{ F_{#1}^{(#2)} }}
\def\cF#1#2{{ {\cal F}_{#1}^{(#2)} }}

\def\R{{\bf R}}
\def\sst#1{{\scriptscriptstyle #1}}
\def\oneone{\rlap 1\mkern4mu{\rm l}}
\def\e7{E_{7(+7)}}
\def\td{\tilde}
\def\wtd{\widetilde}
\def\im{{\rm i}}
\def\bog{Bogomol'nyi\ }
\newcommand{\ho}[1]{$\, ^{#1}$}
\newcommand{\hoch}[1]{$\, ^{#1}$}
\newcommand{\bea}{\begin{eqnarray}}
\newcommand{\eea}{\end{eqnarray}}
\newcommand{\ra}{\rightarrow}
\newcommand{\lra}{\longrightarrow}
\newcommand{\Lra}{\Leftrightarrow}
\newcommand{\ap}{\alpha^\prime}
\newcommand{\bp}{\tilde \beta^\prime}
\newcommand{\cB}{{\cal B}}
\newcommand{\cO}{{\cal O}}
\newcommand{\vecx}{\vec{x}}
\newcommand{\vecy}{\vec{y}}
\newcommand{\vecp}{\vec{p}}
\newcommand{\vecq}{\vec{q}}
\newcommand{\tr}{{\rm tr} }
\newcommand{\Tr}{{\rm Tr} }
\newcommand{\NP}{Nucl. Phys. }

\newcommand{\cL}{{\cal L}}
\newcommand{\cA}{{\cal A}}
\newcommand{\cT}{{\cal T}}
\newcommand{\cR}{{\cal R}}
\newcommand{\cD}{{\cal D}}
\newcommand{\cH}{{\cal H}}

\def\Cb{\bar{C}}

\def\sst#1{{\scriptscriptstyle #1}}
\def\0{{\sst{(0)}}}
\def\1{{\sst{(1)}}}
\def\2{{\sst{(2)}}}
\def\3{{\sst{(3)}}}
\def\4{{\sst{(4)}}}
\def\5{{\sst{(5)}}}
\def\6{{\sst{(6)}}}
\def\7{{\sst{(7)}}}
\def\8{{\sst{(8)}}}
\def\9{{\sst{(9)}}}
\def\p{{\sst{(p)}}}
\def\q{{\sst{(q)}}}
\def\ve{\varepsilon}
\def\vf{\varphi}
\def\F{\Phi}
\def\wg{\wedge}

\def\thb{\bar{\theta}}
\def\Thb{\bar{\Theta}}
\def\barp{\bar{p}}
\def\barq{\bar{q}}
\def\barc{\bar{c}}
\def\bard{\bar{d}}
\def\e{\epsilon}

\def \bi{\bibitem}
\def \la {\label}

\def \l {\lambda}
\def\foot{\footnote}
\def \tl  {{\tilde \l}}
\def \sql {{\sqrt \l}}
\def \adss {$AdS_5 \times S^5$\ }
\newcommand{\rf}[1]{(\ref{#1})}
\def \ov {\over}

\def\th{\theta}
\def\Th{\Theta}
\def\vth{\vartheta}
\def\btheta{{\bar\theta}}
\def\ttheta{{{\tilde\theta}}}
\def\bttheta{{{\bar\ttheta}}}
\def\vth{\vartheta}

\def\ra{\rightarrow}
\def\N{\nabla}
\def\F{{\cal F}}
\def\uM{\underline{M}}
\def\uA{\underline{A}}
\def\uN{\underline{N}}
\def\uP{\underline{P}}
\def\ua{\underline{a}}
\def\ub{\underline{b}}
\def\uc{\underline{c}}
\def\ud{\underline{d}}
\def\ue{\underline{e}}
\def\uf{\underline{f}}
\def\ui{\underline{i}}
\def\uj{\underline{j}}
\def\uk{\underline{k}}
\def\ul{\underline{l}}
\def\ual{\underline{\alpha}}
\def\ube{\underline{\beta}}
\def\um{\underline{m}}
\def\un{\underline{n}}
\def\up{\underline{p}}
\def\uq{\underline{q}}
\def\ur{\underline{r}}
\def\us{\underline{s}}
\def\umu{\underline{\mu}}
\def\unu{\underline{\nu}}
\def\ula{\underline{\l}}
\def\uka{\underline{\k}}
\def\usi{\underline{\s}}
\def\urh{\underline{\r}}
\def\cc{\circ}
\def\eqv{\equiv}

\def\ni{\noindent}

\def\Ep{E^{{}^{(+)}}}
\def\Em{E^{{}^{(-)}}}

\def\Mp{M^{{}^{(+)}}}
\def\Mm{M^{{}^{(-)}}}

\def \ha{{1\ov 2}}

\def\r{\rho}

\def\Y{{\rm Y}}
\def\X{{\rm X}}
\def\tY{\tilde{\rm Y}}
\def\tX{\tilde{\rm X}}
\def\dY{\dot{\rm Y}}
\def\dX{\dot{\rm X}}

\def \J {\mathcal{J}}
\def \del {\partial}

\def\dF{\dot{F}}
\def\dG{\dot{G}}
\def\df{\dot{f}}
\def \E {{\cal E}}
\def \S {{\cal S}}
\def \J {{\cal J}}

\def\ms{\mathcal{S}}
\def\mj{\mathcal{J}}
\def\soj{\fr{\ms}{\mj}}
\def \R {{\bf R}}
\def \om {\omega}
\def \bE {\bar E}
\def \x {{\cal X}}

\def \bi{\bibitem}
\def \la {\label}

\def \l {\lambda}
\def\foot{\footnote}
\def \tl  {{\tilde \l}}
\def \sql {{\sqrt \l}}
\def \adss {$AdS_5 \times S^5$\ }
\def \ov {\over}

\def \varpi {{\rm w}}

\def\thb{\bar{\theta}}
\def\Thb{\bar{\Theta}}
\def\mb{\bar{\m}}
\def\ab{\bar{\a}}
\def\zb{\bar{z}}
\def\psib{\bar{\psi}}
\def\barp{\bar{p}}
\def\barq{\bar{q}}
\def\barc{\bar{c}}
\def\bard{\bar{d}}
\def\e{\epsilon}
\def\wb{\bar{w}}
\def\lb{\bar{\l}}
\def\Jb{\bar{J}}
\def\Nb{\bar{N}}
\def\Zb{\bar{Z}}
\def\pab{\bar{\pa}}

\def\At{\tilde{A}}
\def\Bt{\tilde{B}}
\def\Ct{\tilde{C}}
\def\Dt{\tilde{D}}
\def\Et{\tilde{E}}
\def\Ft{\tilde{F}}
\def\Gt{\tilde{G}}
\def\Ht{\tilde{H}}
\def\Kt{\tilde{K}}
\def\Mt{\tilde{M}}
\def\Nt{\tilde{N}}
\def\Rt{\tilde{R}}
\def\at{\tilde{a}}
\def\bt{\tilde{b}}
\def\ct{\tilde{c}}
\def\dt{\tilde{d}}
\def\et{\tilde{e}}
\def\ft{\tilde{f}}
\def \ztt{\tilde{\z}}
\def\htil{\tilde{h}}
\def\gt{\tilde{g}}
\def\nt{\tilde{n}}
\def\mut{\tilde{\mu}}
\def\nut{\tilde{\nu}}
\def\pht{\tilde{\f}}
\def\vft{\tilde{\vf}}

\def\rht{\tilde{\rho}}

\def\asth{\hat{*}}
\def\phh{\hat{\phi}}

\def\bA{{\bf A}}

\def\ola{\overleftarrow}
\def\ora{\overrightarrow}
\def\alt{\tilde{\a}}

\def\eh{\hat{e}}
\def\eph{\hat{\e}}
\def\ph{\hat{p}}
\def\alh{\hat{\a}}
\def\beh{\hat{\b}}
\def\gah{\hat{\g}}
\def\Fh{\hat{F}}
\def\muh{\hat{\m}}
\def\nuh{\hat{\n}}
\def\thh{\hat{\th}}
\def\rhh{\hat{\r}}
\def\dh{\hat{d}}
\def\ih{\hat{i}}
\def\jh{\hat{j}}
\def\hh{\hat{h}}
\def\nh{\hat{n}}
\def\gh{\hat{g}}
\def\kh{\hat{k}}
\def\deh{\hat{\d}}
\def\wh{\hat{w}}
\def\lah{\hat{\l}}
\def\Ah{\hat{A}}
\def\Kh{\hat{K}}
\def\Nh{\hat{N}}
\def\Rh{\hat{R}}
\def\Ch{\hat{C}}
\def\Omh{\hat{\Omega}}

\def\xh{\hat{x}}

\def\ps{\rlap{\, /}\;\,p }
\def\ks{\rlap{\, /}\;\,k }

\def\gym{g_{YM}}

\def\adot{\dot{a}}
\def\bdot{\dot{b}}
\def\bpa{\bar{\pa}}

\def\pr{\prime}
\def\ssk{\medskip}
\def\clb{\color{blue}}
\def\clr{\color{red}}
\def\clg{\color{green}}

\def\bfA{{\bf A}}
\def\bfB{{\bf B}}
\def\bfK{{\bf K}}
\def\bfU{{\bf U}}
\def\bfX{{\bf X}}
\def\bfY{{\bf Y}}
\def\bfZ{{\bf Z}}
\def\bfg{{\bf g}}
\def\bfn{{\bf n}}

\begin{document}

\overfullrule=0pt
\parskip=2pt
\parindent=12pt
\headheight=0in \headsep=0in \topmargin=0in
\oddsidemargin=0in

\vspace{ -3cm}
\thispagestyle{empty}

 \vspace{0.1cm}

\setcounter{equation}{0}
\setcounter{footnote}{0}
\setcounter{section}{0}

\begin{center}

{\Large\bf One-loop renormalization of a gravity-scalar system}

\vskip 0.8cm

 \vspace{.5cm}

\vspace{0.5cm}
I. Y. Park
\\

\vspace{0.3cm}

\vspace{0.3cm}
{\it Department of Applied Mathematics,
Philander Smith College 
                               \\
Little Rock, AR 72223, USA \\
inyongpark05@gmail.com
}

\end{center}

 \vspace{0.1cm}

\begin{abstract}

Extending the renormalizability proposal of the physical sector of 4D Einstein gravity, we have recently proposed renormalizability of the 3D physical sector of gravity-matter systems. The main goal of the present work is to conduct systematic one-loop renormalization of a gravity-matter system by applying our foliation-based quantization scheme. In this work we explicitly carry out renormalization of a gravity-scalar system with a Higgs-type potential. With the fluctuation part of the scalar field gauged away, the system becomes renormalizable through a metric field redefinition. We use dimensional regularization throughout. One of the salient aspects of our analysis is how the graviton propagator acquires the``mass" term. One-loop calculations lead to renormalization of the cosmological and Newton's constants. We discuss other implications of our results as well: time-varying vacuum energy density and masses of the elementary particles as well as the potential relevance of Neumann boundary condition for black hole information.

\end{abstract}
\newpage

\section{Introduction}

The true degrees of freedom of quantum gravity \cite{DeWitt:1975ys,'tHooft:1974bx,Stelle:1976gc,Antoniadis:1986tu,Weinberg3,Reuter:1996cp,Odintsov:1990qq,Barvinsky:1993zg,Carlip:2001wq,Thiemann:2007zz,Ambjorn:2012jv,Woodard:2014jba,Donoghue:2015hwa} have been evasive, at least to some extent, in spite of an extended search.
The search along the line of canonical quantization in the  past (e.g. \cite{York:1972sj,Moncrief:1989dx,Fischer:1996qg}) was based on the usual 3+1 splitting where the genuine time coordinate was separated out. 
 A different 3+1 splitting in which one of the spatial directions is separated out has been employed in the recent works of \cite{Park:2014tia} and its sequels in which explicit identification of the physical degrees of freedom has been made based on foliation theory: the physical degrees of freedom are the ones associated with a certain hypersurface.

 The relevance of a hypersurface for true degrees of freedom was noted long ago in \cite{York:1972sj} (see also the works of  \cite{Reisenberger:1996pu,Bianchi:2006uf} in the context of loop quantum gravity): a spacelike hypersurface specified up to a conformal factor was identified as the true degrees of freedom based on the fact that the hypersurface can serve as a transverse and traceless spin-two representation of gravity. 
In \cite{Moncrief:1989dx,Fischer:1996qg} Hamiltonian reduction was carried out on a class of 4D manifolds with certain topological restrictions; the reduced Hamiltonian turned out to be the volume of the hypersurface. The configuration reduction approach (see e.g. \cite{Gay-Balmaz:2014ena} for a relatively recent work) makes intensive and extensive use of differential geometry.

Gravity theories do not seem to share the nice property of the non-gravitational gauge theories: only the latter are renormalizable even when the external states in a Feynman diagram are kept offshell. Our recent proposal \cite{Park:2014tia} hinges on the possibility that  non-renormalizability may be overcome once the external states are restricted to a set of states constrained by several physical state conditions. The dynamics of those physical states can be described through a 3D description.\footnote{The reduction to 3D is needed to establish renormalizability at two- and higher- loops. Strictly speaking, it is not needed for the one-loop renormalizability that is the focus of section 3 and 4. The presence of the cosmological constant is important for the renormalizability of the gravity-matter system as will be pointed out later.} 
As reviewed below, the field equations associated with the non-dynamical ADM variables \cite{Arnowitt:1962hi} are imposed as constraints; the solution of the constraints implies reduction of the physical states to 3D in the sense described, e.g., in \cite{Park:2014tia} and \cite{Park:2014noa}. 
We stress that the renormalizability established is valid, unlike that of Yang-Mills theories, only when the external states satisfy certain physical state conditions (thus become ``three-dimensional" and of measure-zero as compared with the offshell states) and thus is {\em not} in conflict with the offshell nonrenormalizability established in the past. In other words, the renormalizability established in \cite{Park:2014tia} and subsequent related works is renormalizability pertaining to the physical states but not that of the offshell Green's functions.\footnote{The renormalizability is in this restricted sense and the covariance is compromised from 4D to 3D at the end. Nevertheless, the formalism does not impose any experimental limitation since it would be only the physical states that one could measure. The limitations listed are the matter of sophistication of the formalism; eventually 4D covariant formalism will of course be more desirable.}

For quantization one needs not only the identification of the true perturbative degrees of freedom but also the equally non-trivial task of explicitly implementing the actual steps of quantization; for instance, the steps of how to integrate various constraints and gauge-fixings into the quantization procedure. In our scheme of quantization, the constraints are Lagrangian analogues of the so-called spatial diffeomorphism and Hamiltonian constraints; they have been dubbed as the shift vector and lapse function constraints, respectively. The strategy for reduction was clearly spelled out in the previous works \cite{Park:2014tia,Park:2014qoa}\footnote{Slightly different procedures of quantization have been presented in subsequent works \cite{Park:2014noa,Park:2015ota,Park:2015xoa}.}: removal of all of the unphysical degrees of freedom from the external states of the Feynman diagrams. The implication of the solution for the shift vector constraint has been highlighted in \cite{Park:2014qoa} and \cite{Park:2015qxa} in the complementary mathematical context of foliation and jet bundle theories: the geometry admits a dual totally geodesic foliation whose associated abelian Lie algebra leads, upon modding out in the jet bundle setup, to the reduction.

As well known, the presence of matter fields makes the divergences worse (see e.g. \cite{Deser:1974cz}). Nevertheless, the quantization scheme of \cite{Park:2014tia} is applicable to a system with a small number of matter fields such as a gravity-scalar system or an Einstein-Maxwell system considered in \cite{Park:2015ybl}: the matter fields can be gauge-fixed to their background values by using part of the diffeomorphism, thus made non-propagating. (For example, if one considers a gravity-scalar system around a flat background one may entirely gauge away the scalar field.) In essence the scalar field materializes into an additional component of the metric and this way the gravity-matter system under consideration becomes ``purely gravitational" (for certain purposes).  
Once one considers the original theory around a more nontrivial background (such as a black hole \cite{Park:2015xoa})\footnote{A qualitative characterization of the kinds of the backgrounds to which the present quantization scheme is applicable can be found in \cite{Park:2015qxa}. It is not clear whether or not the present foliation based quantization scheme will be applicable to a general background (see \cite{Kiriushcheva:2008sf} for a related discussion).}, things become technically more complicated, and explicit demonstration of the renormalization procedure is well worth it. It is the main goal of the present work to illustrate this procedure by taking the case of a gravity-scalar system with a quartic scalar interaction; more specifically we take a real scalar system with the Higgs-type potential for the matter part with a goal to study the implications of a time-dependent background of cosmological relevance.

In quantization of theories with local symmetries the background field (or external field) method (BFM) is an indispensable tool because it allows one to  compute relatively effortlessly the effective action in a covariant form. Although several good reviews on its application to non-gravitational theories are available (see e.g. \cite{Jack:1982hf}), we could not find any review that explicitly addresses two subtle points - which have turned out to be tied with the 4D covariance - that arise in its application to gravity theories. (For the gravity case, see e.g. \cite{Alvarez:1988tb}.) Firstly, unlike a non-gravitational gauge theory in which one can consider the field around a trivial vacuum (meaning $vev=0$), one cannot consider a gravity theory expanded around a zero metric background. This point seems implicitly understood in the literature, however, we find it imperative to keep its implication at the forefront of the background perturbative analysis.\footnote{As a matter of fact, the fact that the conventional way of applying the BFM leads to non-covariance was explicitly stated in \cite{Buchbinder} as we have recently become aware of. (See footnote \ref{odint} for more details.)} The second point is the fact that the trace piece of the fluctuation metric makes the path integral ill-defined. (Strictly speaking, it is a subtlety associated with the path integral itself rather than with the BFM.) Although this fact was observed long ago, we are not aware of any work in which the trace piece was gauge-fixed in the manner discussed in our recent works. 
We illustrate the pathology with several examples including the non-minimal coupling term \cite{Birrell} that has played an important role, e.g., in the Higgs inflation proposal \cite{Bezrukov:2007ep}\cite{Barvinsky:2012zz}\cite{Hamada:2012bp}.

Although dimensional regularization is highly convenient for many purposes it is not as convenient (see \cite{Veltman:1980mj} and \cite{Harada:2000at} for related discussions) for analyzing vacuum diagrams - which are necessary in the present case, e.g., to study renormalization of the Newton's constant - of a massless theory for the reason to be explained. At least for the system considered here there exists a way to get around this inconvenience (the method should apply to any system with a cosmological constant-type of term) and this is one of the salient features of our analysis: the constant piece of the potential is used as a graviton ``mass" term.\footnote{See e.g. \cite{Gabadadze:2003jq} for a discussion of the graviton ``mass" term. {Treating the quadratic term arising from expansion of the cosmological constant term as the graviton ``mass" term is motivated for a purely technically reason just stated; we do not mean that it is the genuine graviton mass term. It appears in the literature \cite{Gazeau:2006uy}, however, that it is perhaps more than a technicality.} {Viewing the quadratic term as the graviton ``mass" term has a further implication, mixing of physical states and unstable state, as we will comment on in section 3.}}
In other words, the constant piece of the potential provides, once expanded, a graviton ``mass" term,  which can be included in the original massless form of the graviton propagator.
Another aspect of the present gravity-matter analysis worth highlighting is the (indicative) way de Sitter spacetime seems to arise from a flat spacetime through the quantum effects.\footnote{Perhaps this observation should not be taken as something entirely new. However, with the present setup of quantization, it is now possible to make this idea more explicit and precise.} If one considers the case where the minimum potential value is chosen to be zero the system admits a flat spacetime with a constant scalar field as a solution. (The case with the nonzero minimum will be discussed as well.) The loop effects generate vacuum energy as will be explicitly shown.

There is potentially an issue of whether or not  the gauge-fixing of the scalar fluctuation makes sense, given that the Higgs quanta have been discovered in nature. The proposed gauge-fixing should not be a good gauge choice if one intends to study various scattering amplitudes of Higgs particles, say, in a curved background. However, it should be suitable for macroscopic cosmological applications. As we will further remark in the conclusion, the proposed gauge-fixing should presumably be viewed as a novel {\em ``scalar-integrating-out"} procedure. It will also be noted later that renormalization with a {\em dynamical} scalar will require only minor modifications.

Although the quantization and quantum corrections may yet be of limited experimental significance, the quantization procedure provides the rationale for gauge-fixing of the scalar, thereby making it non-dynamical in the sense that it no longer runs on the loops of the Feynman diagrams. The non-dynamical scalar field substantially facilitates the renormalization procedure that in turn is crucial for a more concrete establishment of the time-varying vacuum energy. 
In the case of a time-dependent background, we set the analysis of the quantum correction terms aside for now (so that one will not have to worry about the precise values of the coefficients of the finite terms) but focus on its qualitative implications. Our scheme seems to imply time-dependence (and running) of Newton's constant, and reinforce the idea of time-dependent vacuum energy \cite{Peebles:2002gy,Linder:2007wa,Sola:2013gha,Weinberg4}. Although we do not explicitly consider the standard model particles other than the Higgs-type field in this work, our framework is likely to lead, at least as a matter of principle, to time-dependence of various fundamental constants such as masses of standard model particles that become massive through the Higgs mechanism.

\vspace{.3in}
The rest of the paper is organized as follows:

\vspace{.1in}

In section 2 we start by reviewing our main analytical tool, the background field method (BFM). We highlight several potentially subtle points in applying the method in a gravitational setup. Several diagrams are computed by employing the traceless and traceful propagators, and the outcomes are compared: only the traceless propagator leads to 4D covariant results.\footnote{{The necessity of employing the traceless propagator was noted in \cite{Ortin} (ch 3) as I have become recently aware of.}} 
 Afterwards we review our recent proposal on the physical states of gravity theories. The reduction of the physical states to 3D has been established in several different ways; here we give a hopefully more elucidating account of the steps to the reduction with added comments. The reduction will be used, although rather implicitly, in section 3 and 4.
With the reviews of the BFM and physical state reduction, we will be ready for the explicit implementation of the one-loop renormalization of the gravity-scalar system; in section 3 we embark on a one-loop analysis of the scalar-gravity system.
Renormalizability is established with the flat propagator with which the divergence analysis can be easily carried out. 
We gauge away the fluctuation part of the scalar field and set the scalar to its background value.\footnote{The word ``background" is used here (and in some places) to refer to the solution (e.g. $g_{0\m\n}$ in \rf{gshift} below). In other places it refers to the offshell background field (e.g. $\gt_{B\m\n}$ or  $\vf_{B\m\n}$ in \rf{gshift}).}
With the gauge-fixing, the background part of the scalar field provides the scalar background; the fluctuation part will materialize into an additional metric degree of freedom. Although we use the 4D covariant notation, the reduction to 3D will be implicitly conceived as will be explained in more detail.
Since the manifest scalar part is served just as a background the prospect for renormalizability a priori looks better. 
As one of the examples of illustrating the 4D covariance and importance of the quantum effects, we note that the non-minimal coupling between the metric and scalar, which played a crucial role in the Higgs inflation proposal \cite{Bezrukov:2007ep}\cite{Barvinsky:2012zz}\cite{Hamada:2012bp}, is generated by a graviton loop.
We show in dimensional regularization that the loop effects renormalize the cosmological and Newton's constants. 
In section 4 we discuss implications of our results for cosmology. In addition to the time-varying vacuum energy, our framework is likely to imply in a predictable manner time-variations of the masses of elementary particles whose masses originate from the Higgs mechanism, the effects that should be measurable, at least in principle. Our results seem to indicate the possibility that the Higgs field may play the role of the quintessence field as well as the role of the inflaton field.
 In conclusion we summarize and discuss several future directions.
As one of the directions we ponder the possibility that the physical state condition may be at odds with the Dirichlet boundary condition and that the Neumann boundary condition may play a role in solving the black hole information problem. 
Our conventions and useful identities are presented in Appendix A. Several examples of the pathology associated with the traceful propagator can be found in Appendix B.  
In Appendix C we give a coordinate-free version of the analysis in section 2.2 (originally given in section 3.1 of \cite{Park:2014qoa}), the proof that the shift vector constraint leads to the condition for Riemannian foliation.

\section{Review of BFM and reduction}

Before setting out to achieve the main goal of the present work we set the stage by reviewing the background field method (BFM) and reduction of the physical states.    
We extend our recent background field calculations in \cite{Park:2015ota} to the scalar sector. Also, because the reduction of the physical states to 3D has been previously established in several different ways we present here a more coherent review of the reduction with added comments. 
      
Although the BFM was introduced long ago and has been widely used, it appears that there still is room for refinement, especially in applications to the gravitational cases. One of the improvements made in \cite{Park:2014noa} was with regards to the manner in which the problem, noted in \cite{Gibbons:1978ac} and \cite{Mazur:1989by} some time ago, of the ill-defined path integral associated with the trace part of the fluctuation metric was dealt with. We first review that the presence of the trace piece leads to pathology in the perturbation theory: it interferes with the 4D covariance. (As previously stated, the necessity for removal of the trace piece was noted in \cite{Ortin} for a different reason.) This has been demonstrated in \cite{Park:2015ota} in the graviton and ghost sectors. In this work we expand the analysis to the scalar sector; we show that the same pathology is present there as well. In particular, the non-minimal coupling comes out in the covariant form only when the traceless propagator is employed. 

The renormalizability proposal in \cite{Park:2014tia} was based on the observation that the physical states of the gravitational system under consideration lend to a 3D description. In the ADM Hamiltonian formulation one can easily observe the non-dynamism of the shift vector and lapse function. In the ADM Lagrangian formalism these fields can be gauge-fixed and their field equations should be imposed as constraints - the shift vector and lapse function constraints\footnote{Eventually the quantum corrections to these constraints should be considered; we will have more on this in the conclusion.} which are the Lagrangian versions of the momentum and Hamiltonian constraints of the Hamiltonian formalism. The shift vector constraint can be solved as we will review and the solution implies that the geometry should be of the so-called Riemannian foliation \cite{Park:2014qoa}\cite{Park:2015qxa}.

\vspace{.3in}

\ni Our system is the following gravity-scalar action
\bea
S=\fr1{\k^2}\int d^4 x \sqrt{-g}\;(R-2\L)  - \int d^4 x\sqrt{-g}\; \Big(\fr12g^{\m\n}\pa_\m\z \pa_\n \z +V\Big)
\la{grv-sclr}
\eea
where $\k^2= 16\pi G$ with $G$ Newton's constant; $\k^2$ will be suppressed in some places.
The potential $V$ is given by
\bea
V=\fr{\l}{4}\Big(\z^2+\fr{1}{\l} \m^2\Big)^2  \la{mzpot}
\eea
where $\l$ is the scalar coupling and $\m^2$ is the mass parameter.
Note that the cosmological constant $\L$ serves as the minimum value of the potential. The presence of the cosmological constant is important as we will see below. (This point has been stressed in \cite{Park:2016fxc}.) A shift of the scalar potential by a constant is immaterial in a flat spacetime quantum field theory. The same is not true, however, in a curved case; we have introduced an arbitrary cosmological constant term to keep things at a more general level.

\subsection{BFM and 4D covariance}

In section 3 the renormalization of \rf{grv-sclr} will be carried out with the BFM. As pointed out in \cite{Park:2015ota}, care is required to avoid the pathology caused by the trace mode of the fluctuation metric. In this subsection we start with several warm-up calculations. 

We consider the {\em conventional} (except the removal of the trace piece of the fluctuation metric) setup for illustrative purposes: not just the internal momenta but also the external momenta will be kept offshell. In other words we do not employ the reduction scheme in section 2.1; this is to demonstrate the pathology in a setup as conventional as possible. By the same token we do not gauge-fix the scalar. That is, the objects computed are the offshell Green's functions (other than the removal of the metric trace mode). We contrast the amplitudes obtained by employing the traceful propagator against those calculated by employing the traceless propagator, and show that only the traceless propagator leads to the anticipated covariant results. 
Let us start with the gravity sector:  
\bea
S=\fr1{\k^2}\int d^4 x \sqrt{-g}\;R  
\la{unsplit}
\eea
By shifting the metric according to\footnote{This potentially peculiar-looking shift is to deal with the non-covariance issue stated in \cite{Buchbinder}.  {In section 3.6.1 of \cite{Buchbinder}, the technique how to compute the counter-terms is commented. It is the standard method based on the usual shift $g_{\mu\n}=\eta_{\mu\nu}+h_{\mu\nu}$. At the end of that section, it is written, ``However, explicit general coordinate invariance is not manifest in this case. It is necessary to do some work to write counterterms in general covariant form." The present analysis with the double-shift corresponds to ``doing some work to write counterterms in general covariant form."} \label{odint} {(Interestingly, it appears that the essentially same shift had been employed in \cite{Antoniadis:1995fc} as I have come to know very recently.)}}
\bea
g_{\m\n}\equiv  h_{\m\n}+\tilde{g}_{{}_B\m\n}\quad \mbox{where}\quad \tilde{g}_{{}_B\m\n}\equiv \vf_{{}_B\m\n}+g_{0\m\n} \la{gshift}
\eea
one obtains \cite{Goroff:1985th}\cite{Park:2015ota}\footnote{The overall factor $\fr1{\k'^2}$ where $\k'^2 \equiv 2\k^2 $ has been suppressed.}
\bea
\hspace{1in} &&\cL = \sqrt{-\gt}\,\Big( -\fr12\tilde{\N}_\g h^{\a\b}\tilde{\N}^\g h_{\a\b}+\fr14 \tilde{\N}_\g h^{\a}_\a \tilde{\N}^\g h^{\b}_\b  \la{gravcubcov}
\\
&&\hspace{-1in}+h_{\a\b}h_{\g\d}\Rt^{\a\g\b\d}-h_{\a\b}h^{\b}{}_\g \Rt^{\k\a\g}{}_{\k}
{ -}h^{\a}{}_{\a}h_{\b\g}\Rt^{\b\g}-\fr12 h^{\a\b}h_{\a\b}\Rt
+\fr14  h^{\a}_\a  h^{\b}_\b \Rt +\cdots\Big) \nn
\eea
where $(\cdots)$ stands for the terms irrelevant to the present one-loop counterterm computation. The fields with a tilde are the background quantities constructed out of $\tilde{g}_{{}_B\m\n}$. In \cite{Park:2014noa, Park:2015ota} the counterterms obtained with the shift specified in \rf{gshift} were explicitly demonstrated to cancel the divergences in the one-loop Feynman diagrams.
Similarly, the ghost action can be expanded\footnote{The sign error here affected several equations below. They must be further corrected. (They have been corrected in \cite{Park:2015ota}.)}: 
\bea
{\cal L}_{\mbox{gh}}&=&  -\tilde{\N}^\n \Cb^\m \tilde{\N}_\n C_\m  { +}\Rt_{\m\n}\bar{C}^\m C^\n +\cdots 
\eea
Here and in \rf{gravcubcov} only the one-loop pertinent terms have been kept. 
The divergence part can be computed by setting $g_{0\m\n}$ to be a flat metric, $g_{0\m\n}=\eta_{\m\n}$. This is because the divergence must come from a short distance. For the finite parts (whose precise evaluation will not be pursued in this work) one should use the propagator obtained by using the actual background of the solution metric $g_{0\m\n}$ (or $\tilde{g}_{{}_B\m\n}$, more precisely).

\subsubsection*{ghost loop}

To see the pathology caused by the trace piece of the fluctuation metric let us first consider the ghost kinetic term
\bea
{\cal L}_{\mbox{gh, kin}}&=&  -\sqrt{-\gt}\;\tilde{\N}^\n \Cb^\m \tilde{\N}_\n C_\m
\la{ghkin}
\eea
If all is well this kinetic term is expected to yield covariant forms of the counterterms once the ghost field is integrated out. As explicitly shown at one-loop in \cite{Park:2015ota}, use of the traceless propagator has confirmed this expectation whereas the traceful propagator has contradicted it. 
Let us expand the metric $\gt_{\m\n}$ around a flat spacetime:
\bea
\gt_{\m\n}\equiv \vf_{B\m\n}+\eta_{\m\n}
\eea
Below we will often omit the letter `$B$' from $\vf_{{}_B\m\n}$. The ghost kinetic action \rf{ghkin} takes
\bea
{\cal L}_{\mbox{gh}}&=& -\Big[
\pa^\m \bar{C}^\n \pa_\m {C}_\n+\fr{1}{2}\vf \pa^\m \bar{C}^\n \pa_\m {C}_\n
-\tilde{\G}^\l_{\m\n}( \pa^\m \bar{C}^\n { C_\l}-  \pa^\m {C}^\n \bar{C}_\l  )\nn\\
&& -(\eta^{\n\b}\vf^{\m\a}+\eta^{\m\a}\vf^{\n\b})\pa_\b \bar{C}_\a \pa_\n {C}_\m
+\Rt_{\m\n}\Cb^\m C^\n\Big]  \la{ghkinexp}
\eea
Briefly, the result of the computation is as follows. (More details can be found in \cite{Park:2015ota}.) 
The following is the correlator to be computed to obtain the counterterms for the graviton two-point amplitude with the ghost loop: 
\bea
\hspace{-.4in}\fr{i^2}2<\Big[\int \fr{1}{2}\vf \pa^\m \bar{C}^\n \pa_\m {C}_\n
-\tilde{\G}^\l_{\m\n}( \pa^\m \bar{C}^\n { C_\l}-  \pa^\m {C}^\n \bar{C}_\l    )
-(\eta^{\n\b}\vf^{\m\a}+\eta^{\m\a}\vf^{\n\b})\pa_\b \bar{C}_\a \pa_\n {C}_\m  +\Rt_{\m\n}\Cb^\m C^\n
\Big]^2>\nn\\
\eea
This correlator leads to the following counterterms: 
\bea
\D \cL &=& -\fr12 \fr{\G(\e)}{(4\pi)^2}\Big[ \;{ -\fr{2}{15}}\pa^2\vf_{\m\n}\pa^2 \vf^{\m\n} { +\fr{4}{15}}\pa^2 \vf^{\a\k}\pa_\k \pa_\s \vf_\a^\s
{ -\fr{1}{30}}(\pa_\a \pa_\b \vf^{\a\b})^2
\Big]  \nn\\
&=& -\fr1{30} \fr{\G(\e)}{(4\pi)^2}\Big[{ -8}\Rt_{\a\b}\Rt^{\a\b}+{ \fr72}\Rt^2\Big] \la{ctrghpartI2}
\eea
where the parameter $\ve$ is related to the total spacetime dimension $D$ by
\bea
D=4-2\ve
\eea
The relations in \rf{covctr} - which of course are valid to the second order of the fields - have been used to derive the second equality in \rf{ctrghpartI2};  the trace piece has also been set to zero, $\vf\equiv \vf^\a_\a=0$.\footnote{The constraint associated with the fixing of the trace piece of the fluctuation can be identified with one of the components of the nonlinear de Donder gauge (or the generalization thereof) \cite{Park:2015xoa}.}

\subsubsection*{graviton loop}

We now turn to the graviton sector with the graviton loop and observe the same phenomenon: only the traceless propagator leads to covariant results. By expanding \rf{gravcubcov} one gets
\bea
\cL&=& -\fr12 {\pa}_\g h^{\a\b}{\pa}^\g h_{\a\b}+\fr14 {\pa}_\g h^{\a}_\a {\pa}^\g h^{\b}_\b  \nn\\
&& + \cL_{V_I}+\cL_{V_{II}}+\cL_{V_{III}}   \la{eawv}
\eea
where
\bea
\cL_{V_I} &=&   \Big(2\eta^{\b\b'}\tilde{\G}^{\a' \g\a}- \eta^{\a\b}\tilde{\G}^{\a' \g\b'}\Big)\pa_\g h_{\a\b}\, h_{\a'\b'}  \nn\\
\cL_{V_{II}} &=& \Big[\fr12(\eta^{\a\a'}\eta^{\b\b'}\vf^{\g\g'}+\eta^{\b\b'}\eta^{\g\g'}\vf^{\a\a'}
+\eta^{\a\a'}\eta^{\g\g'}\vf^{\b\b'})\nn\\
&&-\fr14 \vf\, \eta^{\a\a'}\eta^{\b\b'}\eta^{\g\g'}-\fr12 \eta^{\g\g'}\eta^{\a'\b'}\vf^{\a\b}  \nn\\
&&+\fr14 (-\vf^{\g\g'}+\fr12 \vf \eta^{\g\g'})\eta^{\a\b}\eta^{\a'\b'}
\Big] \pa_\g h_{\a\b}\, \pa_{\g'}h_{\a'\b'}  \la{lv12}
\eea
These two vertices come from the first line of \rf{gravcubcov}. The distinction between $\cL_{V_I} $ and $\cL_{V_{II}}$ has been made for convenience in Mathematica coding. The vertex $\cL_{V_{III}}$ is just the third line of the same equation:
\bea
\hspace{-.2in}\cL_{V_{III}} = \sqrt{-\gt}\Big( h_{\a\b}h_{\g\d}\Rt^{\a\g\b\d}-h_{\a\b}h^{\b}{}_\g \Rt^{\k\a\g}{}_{\k}
+h^{\a}{}_{\a}h_{\b\g}\Rt^{\b\g}-\fr12 h^{\a\b}h_{\a\b}\Rt
+\fr14  h^{\a}_\a  h^{\b}_\b \Rt \Big)  \la{gver}  \nn\\
\eea
One should consider the following correlator for the counterterms against the complete one-loop two-point amplitude:
\bea
&&\fr{i^2}{2}< \Big[(\cL_{V_I}+\cL_{V_{II}}+\cL_{V_{III}})\Big]^2 >\\
&=&\fr{i^2}{2}\int\int\Big(< (\cL_{V_I}+\cL_{V_{II}})^2 >+2< (\cL_{V_I}+\cL_{V_{II}})\cL_{V_{III}} >
+< \cL_{V_{III}}^2 > \Big) \la{gcor} \nn
\eea 
The third term is guaranteed to be covariant. (This would of course be true even if the traceful propagator were used.)
For the other two correlators the computation employing the traceful propagator is presented in Appendix B; it leads to non-covariant results.
With the traceless propagator the third correlator can be written as
\bea
&&\hspace{-.7in}-\fr12\int\int<\Big(h_{\a\b}h_{\g\d}\Rt^{\a\g\b\d}-h_{\a\b}h^{\b}{}_\g \Rt^{\k\a\g}{}_{\k}
+h^{\a}{}_{\a}h_{\b\g}\Rt^{\b\g}-\fr12 h^{\a\b}h_{\a\b}\Rt
+\fr14  h^{\a}_\a  h^{\b}_\b \Rt \Big)^2> \nn\\
\hspace{.2in}&=&-\fr12\int\int<\Big(h_{\a\b}h_{\g\d}\Rt^{\a\g\b\d}-h_{\a\b}h^{\b}{}_\g \Rt^{\k\a\g}{}_{\k}
-\fr12 h^{\a\b}h_{\a\b}\Rt
\Big)^2> 
\eea
where the equality is due to the use of the traceless propagator: the trace-piece containing terms have been removed.
After some algebra, one gets
\bea
-< \Big[\int\cL_{V_{III}}\Big]^2 > \;\Rightarrow\D \cL  &=& \fr{\G(\e)}{(4\pi)^2}\Big[-\fr54\Rt^2-\fr34 \Rt_{\m\n\r\s}\Rt^{\m\n\r\s} +\fr32\Rt_{\m\n}\Rt^{\m\n}\Big]   \nn\\
                                       &=& -\fr12\fr{\G(\e)}{(4\pi)^2}\Big[-3\Rt_{\m\n}\Rt^{\m\n}-\Rt^2\Big]  \la{tgravi}
\eea
where in the thrid equality the identity \rf{Riemannsqid} has been used.
With the traceless propagator the first correlator on the right-hand side of \rf{gcor} turns out to be
\bea
\hspace{-.3in} -\fr12 < \Big[\int(\cL_{V_I}+\cL_{V_{II}})\Big]^2 > \;\Rightarrow\D \cL  &=&\Big[\fr{7}{20} (\pa^\a\pa^\b \vf_{\a\b})^2-\fr{37}{40}\pa^\a \pa^\r \vf_{\a\b} \pa^2 \vf^\b_\r
   +\fr{37}{80} \pa^2 \vf_{\a\b}  \pa^2 \vf^{\a\b}\Big] \nn\\
   &=& -\fr12 \fr{\G(\ve)}{(4\pi)^2}\Big[-\fr{23}{40}\Rt^2+\fr{37}{20} \Rt_{\a\b}\Rt^{\a\b}\Big] \la{diag}
\eea
The result of the second correlator is
\bea
-< \int\int(\cL_{V_I}+\cL_{V_{II}})\cL_{V_{III}} > \;\Rightarrow\D \cL&= &- \fr{\G(\ve)}{(4\pi)^2}\Big[ \fr12(\pa^\a \pa^\b \vf_{\a\b})\Rt \Big]  
                                                        = -\fr12\fr{\G(\ve)}{(4\pi)^2} \Rt^2   \la{cross} \nn\\
\eea

\subsubsection*{scalar loop}

Our final example of illustrating the pathology is the diagram given in Fig. 1. As a matter of fact there exists the second diagram which will be considered in the next section that generates the counterterm of the non-minimal coupling. We have chosen the diagram above to demonstrate the pathology with a minimum amount of computation.
	Since the form of the non-minimal coupling was not present in the starting Lagrangian, generation of such a counterterm is  incompatible with renormalizability. In this subsection we set the renormalizability issue aside and focus only on the covariance issue to demonstrate the pathology of the trace piece of the metric. In the next section we will employ the setup suitable for establishing the renormalizability wherein only the second diagram remains relevant. We show that use of the traceless propagator yields the counterterm of the well-known non-minimal coupling \cite{Birrell}. 
\bea
&&\begin{tikzpicture}[line width=1 pt, scale=1]
\draw (-140:1)--(0,0);
\draw (140:1)--(0,0);
\draw (.52,0) circle (.49cm);	

\node at (-140:1.2) {$\z_B$};
\node at (140:1.3) {$\z_B$};

\begin{scope}[shift={(1,0)}]
\draw[vector] (0:1)--(0,0);
\node at (0:1.4) {$\gt_B$};
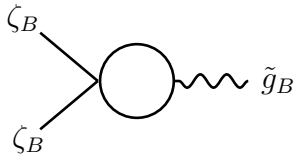
\node at (70:-1.4) {Figure 1: non-minimal coupling (scalar loop)};	
\end{scope}
\end{tikzpicture} \nn
\eea
Let us shift the scalar field
\bea
\z\rightarrow \z_B+{\z} 
\eea
As in the previous examples the traceful propagator does not lead to a covariant result.
The vertex involving the graviton comes from expansion of the scalar kinetic term which yields
\bea
&&\hspace{1.3in} -\fr12\sqrt{-g}\; g^{\m\n}\pa_\m\z \pa_\n \z \rightarrow \\
 && \hspace{-.4in} -\fr12\Big[{ \gt}_B^{\m\n}-h^{\m\n}+\fr12 h { \gt}_B^{\m\n}+h^{\m\r}h_{\r}^\n+\fr18 { \gt}_B^{\m\n}(h^2-2h_{\r\s}h^{\r\s})-\fr12 hh^{\m\n}+\cdots \Big] \pa_\m{ \z}  \pa_\n { \z}       \la{sgsector}  \nn
\eea
Among these terms the vertex relevant for the diagram is the second term on the right-hand side of
\bea
-\fr12 \gt_B^{\m\n} \pa_\m { \z}  \pa_\n { \z} = -\fr12 g_0^{\m\n} \pa_\m { \z}  \pa_\n { \z} +\fr12 \vf_B^{\m\n} \pa_\m { \z}  \pa_\n { \z} 
\eea 
The other vertex in the diagram is the quartic scalar self-coupling. With these two vertices inserted the counterterm turns out to be 
\bea
\fr{i^2}{2} \Big< \int \vf^{\m\n}\pa_\m\z \pa_\n \z\;\int \fr{6\l}{{ 4}}  \z^2\z_B^2\Big> \;\Rightarrow\D \cL 
= \fr{\G(\ve)}{(4\pi)^2}\,\fr\l{{ 4}}\pa_\m\pa_\n \vf^{\m\n} \z_B^2= \fr{\G(\ve)}{(4\pi)^2}\fr\l{{4}}\;R \,\z_B^2 \nn\\
\eea
in the leading order of the metric field.

\subsection{reduction of physical states}

Another central element of the proposal in \cite{Park:2014tia} is the reduction of the physical states.
(The coordinate-free version of the analysis of this subsection can be found in Appendix C.) 
Consider a background that admits 3+1 foliation; split the coordinates into
\bea
x^\m\equiv (y^m,x^3) \la{coord}
\eea
Introducing the lapse function $n$ and shift vector $N_{m}$, the 3+1 form of the 4D metric and its inverse can be written \cite{Arnowitt:1962hi}\cite{Poisson}
\bea
g_{\m\n}=\left(
\begin{array}{cc}
	h_{mn} & N_{ m} \\
	&\\
	N_{ n} &  n^2+h^{mn}N_{m} N_{ n} 
\end{array}
\right)\quad,\quad
g^{\m\n}=\left(
\begin{array}{cc}
	h^{mn}+\fr1{n^2}N^m N^n & -\fr1{n^2}N^{ m} \\
	&\\
	-\fr1{n^2}N^{ n} &  \fr1{n^2} 
\end{array}
\right)   \nn\\
\eea
The action in terms of the ADM variables is
\bea
S=\int d^4 x\;n\sqrt{-\g} \left(R^{(3)}+K^2-K_{mn}K^{mn}\right)
\la{1p3act}
\eea
Here the second fundamental form $K_{mn}$ and its trace $K$ are given by
\be
K_{mn}=\fr1{2n}\left(\mathscr{L}_{\pa_{3}} h_{mn}-{D}_m N_{n}
-{D}_n N_{ m} \right),\qquad K=h^{mn}K_{mn}.
\la{K4defqq}
\ee
with $\mathscr{L}_{\pa_{3}}$ the Lie derivative along the vector field $\pa_{x^3}$ and $D_m$ the 3D covariant derivative constructed out of $h_{mn}$. 
As well known the Hamiltonian of the system reveals that the lapse function and shift vector are non-dynamical; their field equations should be imposed as constraints. In the Hamiltonian formalism the constraints are called the Hamiltonian and momentum constraints. In this work they will be called the lapse function and shift vector constraints, respectively.
The shift vector field equation is
\bea
{D}_m (K^{mn}-h^{mn} K)=0  \la{Ncon}
\eea
The shift vector can be gauged away by using the 3D residual gauge symmetry \cite{Park:2014noa}
\bea
N_{m}=0,   \la{Nazero}
\eea
Substitution of $N_{m}=0$ into \rf{Ncon} leads to
\bea
{D}^m \left[\fr{1}{n}\Big(\mathscr{L}_{\pa_3} h_{mn}
-h_{mn}h^{pq}\mathscr{L}_{\pa_3} h_{pq}\Big)\right]=0  \la{mtmconstr}
\eea
which in turn implies \cite{Park:2014qoa}
\bea
\pa_m n=0 \la{constronn}
\eea
This follows from the fact that the covariant derivative in \rf{mtmconstr} yields zero when it acts on the terms 
other than $\fr1{n}$. To see this, consider for instance the first term inside the parenthesis in \rf{mtmconstr}:  
\bea
D_a \mathscr{L}_{\pa_{3}} h_{bc}=e_a^\a  \N_\a \mathscr{L}_{\pa_{3}} h_{bc}
\la{liecocompresent}
\eea
Applying the component version of the commutator of the Lie derivative and covariant derivatives, \rf{liecocom},  to the present case, one sees that the Lie derivative commutes with the covariant derivative: 
\bea
[\mathscr{L}_{\pa_3}, \N_{\a} ]=\N_{[\pa_{3},\pa_\a]}=\N_{0}=0
\eea
where the last equality is due to the linearity of $\N$ in its index.
With this the right-hand side of \rf{liecocompresent} can be written as
\bea
=e_a^\a   \mathscr{L}_{\pa_{3}} \N_\a h_{bc}
\eea
As stated in e.g., \cite{Poisson} (which can be easily checked), one has $\mathscr{L}_{\pa_{3}} e_a^\a=0 $. With this the right-hand side becomes
\bea
= \mathscr{L}_{\pa_{3}} e_a^\a  \N_\a h_{bc} =\mathscr{L}_{\pa_{3}}   D_a h_{bc}=0
\eea
where the last equality follows from the 3D metric compatibility of the 3D 
covariant derivative.
The reduction to 3D is induced by the lapse function constraint as follows. Since all of the gauge freedom is used, the number of the physical components is fixed, prior to considering the lapse constraint, to be two. Any additional constraint that would reduce independent components, therefore, should not arise from the lapse constraint. The only way of ensuring this is the reduction in the coordinate dependence, which will make the $K_{mn}K^{mn}$ term "identically" vanish. At this point one can legitimately go to the Euclidean space by a Wick rotation; it follows \cite{Park:2014tia,Park:2015xoa} that 
\bea
K_{mn}=0
\eea
As a matter of fact the reduction takes place rather generically even in the presence of matter \cite{Park:2015ybl} due to the fact that the lapse function constraint - which is a {\em constraint} - becomes identical to the ``Hamiltonian''
itself after the shift vector constraint is solved and the gauge-fixing is explicitly enforced.

\section{Renormalization of a gravity-scalar system}

In this section we carry out systematic one-loop renormalization (see e.g. \cite{Sterman} for a review of renormalization in quantum field theory). There are several ingredients in our quantization proposal that make a difference with regards to renormalizability. In the case of pure gravity the reduction of the physical states to 3D is the only condition needed to establish the renormalizability. Once matter fields are added, several additional ingredients are required. Among other things the reduction mechanism itself should be generalized \cite{Park:2015ybl}. Gauging-away of the fluctuation part of the matter field(s) (the scalar for the present case) has been proposed in the cases where the number of matter components is small, as in the present case. For a system with more matter fields this method cannot be applied ; one must directly deal with the dynamical matter fields. For such cases renormalizability remains speculative  \cite{Park:2015ybl} and more work will be required for a definite answer. (See, however, the comments on the recent progress in the conclusion.)

Although gauge-fixing of the fluctuation part of the scalar field, which will be reviewed below,  is a relatively simple step, it greatly facilitates the renormalization procedure in the following sense. Only the metric remains dynamical: all of the scalar field factors can be viewed as backgrounds. The gauge-fixing has the following implication for the non-minimal coupling. The minimal coupling is not a term initially present in our starting action, \rf{grv-sclr}, thus it is one of the terms that would make the system non-renormalizable. In the present setup, however, the non-minimal coupling can be viewed as a ``shift" in the Newton's constant. In other words, all the terms containing a single factor of $R$ and differing by their scalar-field containing coefficients can eventually be grouped into the form $R (\cdots)$ where $(\cdots)$ represents various constants and scalar-involving factors. Furthermore, none of these terms is to be taken as additional {\em new} vertices: the dynamical part of all those vertices is still just $R$ with the scalar field serving as the background but not as a propagating field.

As in section 2 we employ the flat space propagator to compute one-loop counterterms. (The finite parts and sub-leading divergences will be needed for renormalization conditions; for those one should employ the propagator in the actual background geometry.)  
Use of the flat space propagator has an additional meaning in the case of $\L=0$ for the following reason. If the minimum potential value is chosen to be zero then the system has a flat spacetime with a constant scalar field as a solution\footnote{In other words if the potential is shifted by an arbitrary constant $\L\neq 0$ in \rf{grv-sclr} the system will not admit a flat spacetime as a solution.} since the potential term in the metric field equation is evaluated at the constant scalar field that minimizes it. Therefore the analysis can be viewed as the perturbative renormalization around one of the {\em genuine} backgrounds of the system.

Although dimensional regularization is highly convenient in general, the following identity makes it less suited for vacuum diagrams of a massless theory,
\bea
\int d^D k \fr1{(k^2)^\w}=0
\eea
where $\w$ is any number. The divergences that would otherwise renormalize the cosmological and Newton's constants vanish due to this identity. 
One can avoid this\footnote{See e.g. \cite{Veltman:1980mj} and \cite{Harada:2000at} for related discussions.} by introducing a ``mass" term to the graviton propagator in the following manner. The constant piece of the scalar potential provides the following cosmological constant type term:
\bea
-\int \sqrt{-g}\; \Big(\fr{\m^4}{4\l}\Big) \la{pcam}
\eea
Once expanded this term yields a graviton ``mass" term as given in \rf{gmass} below.\footnote{This ``mass" term is not unique: the cosmological constant term $\L$ in \rf{grv-sclrq3} could be added as well. The two choices should be just two different (but equivalent) approaches. We choose \rf{pcam} as the mass term for an occasion where $\L=0$ may be considered.} (See e.g. \cite{Gabadadze:2003jq} for a discussion of the graviton mass terms.)
As we will see shortly, the ``mass" term preserves the tensor structure of the original graviton propagator; it only changes the propagator by adding the mass square next to the momentum square. One can carry out the divergence analysis with the ``massive" propagator. We will see that the cosmological constant term and Einstein-Hilbert term become renormalized through the loop effects.

{Although the use of the ``massive" propagator is technically motivated, it may have deeper physical meanings. As stated in footnote 6, one implication is the possibility proposed in \cite{Gazeau:2006uy}. Another implication is for the renormalization procedure. In the Standard Model Higgs physics, one explicitly expands the action around the true vacuum in conducting the loop analyses. However, in the refined application of the BFM that we have been using, the expansion is more implicit\footnote{See \cite{Carvalho:2013wsa} and \cite{Pelissetto:2015yha} for the recent examples of the renormalization analyses without explicit expansion around the true vacuum.} for the sake of maintaining the covariance, and for this reason inclusion of the cosmological constant in the graviton propagator implies ``mixing" between the graviton field and unstable Higgs field. Often the system under consideration imposes such mixing for one reason or another. As a matter of fact there have been recent studies \cite{Pius:2014iaa} on renormalization procedure of massive string modes that display such mixing.  
}

We quote the gravity-scalar action \rf{grv-sclr} here now in terms of the renormalized quantities (indicated by the subscript $r$):
\bea
S=\fr1{\k_r^2}\int d^4 x \sqrt{-g_r}\;(R_r-2\L_r)  - \int d^4 x\sqrt{-g_r}\; \Big(\fr12g_r^{\m\n}\pa_\m\z_r \pa_\n \z_r +V_r\Big)
\la{grv-sclrq3}
\eea
where 
\bea
V_r=\fr{\l_r}{4}\Big(\z_r^2+\fr{1}{\l_r} \m_r^2\Big)^2
\eea
Below we consider graphs with up to two $g$- or four $\z$- external legs; they should be sufficient for illustrating the renormalization procedure.

\subsection{loop effects in gravity sector}

The aforementioned inconvenient feature of dimensional regularization can be avoided for the present system by including in the original massless propagator the ``mass" term originating from the scalar potential. (Recall that the loop-induced cosmological constant-type of terms were not generated in the early works because the massless graviton propagator was used with dimensional regularization.) With the new ``massive" propagator the one-loop correction for the cosmological term does not vanish. Similarly, there will be renormalization of the Newton's constant.

The analysis required to show the cosmological constant renormalization is a curved space generalization of the standard calculation for which we refer, to be definite, to section 16.2 of \cite{Weinberg2}. When expanded the quadratic order of the constant piece of the scalar potential yields
\bea
\fr14 m^2\int  (h^2-2 h_{\m\n}h^{\m\n}) \la{gmassterm}
\eea
where
\bea
\ m^2 \equiv  \fr18 \fr{\k'^2 \m^4}{\l} \la{amass}
\la{gmass}
\eea
with
\bea
\k'^2 \equiv 2\k^2 
\eea
The term \rf{gmassterm} can be combined with the original massless propagator of the graviton; the first line of \rf{eawv} becomes, after multiplying an overall factor two to be consistent with the convention of \rf{eawv},
\bea
\cL&=& -\fr12 {\pa}_\g h^{\a\b}{\pa}^\g h_{\a\b}+\fr14 {\pa}_\g h^{\a}_\a {\pa}^\g h^{\b}_\b  
    + m^2 \Big(-\fr12 h_{\m\n}h^{\m\n}+\fr14 h^2 \Big) \la{eawm}
\eea
The propagator then is given by
\bea 
<\f_{\m\n}(x_1)\f_{\r\s}(x_2)>&=& P_{\m\n\r\s} \;\k'^2\int \fr{d^4k}{(2\pi)^4}\fr{e^{ik\cdot (x_1-x_2)}}{i( k^2+m^2)} 
\eea
where $P_{\m\n\r\s}$ for the traceless propagator takes the same form as before:
\bea
P_{\m\n\r\s}\equiv \fr12(\eta_{\m\r}\eta_{\n\s}+\eta_{\m\s}\eta_{\n\r}- \fr12\eta_{\m\n}\eta_{\r\s})
\eea
The contribution leading to the cosmological constant renormalization comes from
\bea 
\int \prod_x dh_{\k_1\k_2}\;e^{\fr{i}2 \int \sqrt{\gt} \;h^{\a\b}({\pa}^2-m^2) h_{\a\b}  }
\eea
Then by following the analysis given in \cite{Weinberg2} one obtains a constant term therein denoted by ``$\mathscr{I}$." Let us denote by $\mathscr{I}_{div}$ the divergent part of the analogous quantity in our case. In the curved spacetime context the constant $\mathscr{I}$ must be, due to the 4D covariance, nothing but the coefficient of the cosmological constant-type term. The renormalization of the Newton's constant can be similarly established. Let us consider the following graviton vertices:
\bea
\cL_{V_{III}} = \sqrt{-\gt}\Big( h_{\a\b}h_{\g\d}\Rt^{\a\g\b\d}-h_{\a\b}h^{\b}{}_\g \Rt^{\k\a\g}{}_{\k}-\fr12 h^{\a\b}h_{\a\b}\Rt
 \Big)  \la{gvertl}  
\eea
By self-contracting the $h_{\m\n}$ factors one gets
\bea
&\rightarrow & \D \cL={3} \fr{\G(\ve)}{(4\pi)^2}m^2 \sqrt{-\gt}\;\Rt
\la{ncrenc1}
\eea

\subsection{loop effects in scalar-involving sector}

As reviewed in section 2, the pure gravity system reduces to 3D once the shift vector constraint is enforced. The reduction has been extended in \cite{Park:2015ybl} to certain gravity-matter systems including the present one.
Once a matter field is added one encounters various matter-containing vertices and proliferation of the required counterterms as the number of the loops increases; for the gravity-scalar system it has been proposed to gauge away the fluctuation part of the scalar field by utilizing the residual diffeomorphism with 3D coordinate dependence.\footnote{
	The number of the gauge parameters and the corresponding fixings go as follows \cite{Park:2015ybl}: there are four bulk gauge parameters and corresponding four measure-zero 3D residual paramters.
	The three bulk gauge parameters can be used for the $\m = 0, 1, 2$ components
	of the de Donder gauge. The three measure-zero residual 3D gauge parameters (see the analysis of the residual symmetry in \cite{Park:2014noa}) can be used for the shift vector fixing. Lastly one remaining parameter with the 3D coordinate dependence {can be used} for gauge-fixing of the scalar since the scalar field at this point becomes reduced to 3D. (The metric-trace fixing can be effectively executed by employing the traceless propagator and the lapse function is fixed automatically from these fixings.)
	As commented in our previous works (see e.g. \cite{Park:2015ybl}), it is possible to adopt a slightly different gauge-fixing procedure. For instance, one may use the 4D diffeomorphism to gauge-fix the scalar and shift vector prior to imposing the de Donder gauge. In either of these cases, one should rely on the reduction at the end for complete gauge-fixing.} With this gauge-fixing, the background part of the scalar field provides the background and the fluctuation part materializes into a metric degree of freedom. Since the manifest scalar part (i.e., the remaining part) serves just as a background, renormalizability can be established in the sense to be discussed. We start by reviewing the gauge-fixing procedure of the scalar field, which is well known in the field of scalar-driven inflation, followed by the loop analysis, which will set the stage for establishing renormalizability in section 3.3.

\subsubsection{gauge-fixing of scalar}

Consider a generic scalar field $\s(x)$ coupled with gravity:
\bea
\s(x):\quad \mbox{a generic scalar field}
\eea
Its infinitesimal and finite diffeomorphism transformations, respectively, are 
\bea
\s'(x)=\s(x)+\xi^\m \pa_\m \s(x)  \la{gr}
\eea
and (see, e.g., \cite{GPP})
\bea
\s'(x)=e^{\mathscr{L}_\xi}\s(x)
\eea
where $\mathscr{L}_\xi$ denotes the Lie derivative along the vector field $\xi$. Let us first consider the gauge-fixing of the scalar in a generic metric background. In the background field method one shifts the original field to
\bea
\s\equiv \s_0+\hat{\s}
\eea
where $\s_0$ denotes the classical solution and $\hat{\s}$ the fluctuation. For the reason explained in the earlier works  \cite{Park:2014noa, Park:2015ota} let us introduce another shift $\hat{\s}\equiv \bar{\s}+\s_f$ and consider
\bea
\s\equiv \s_B+\s_f \quad,\quad  \s_B \equiv  \s_0+{\bar{\s}}
\eea
where $\s_f$ is taken as the fluctuation field and $\bar{\s}$ is analogous to $\vf_{B\m\n}$ in \rf{gshift}.
When computing Feynman diagrams the $\s_B$ fields are placed in the external lines.
Coming back to the present case, let us define
\bea
\z\equiv \z_B+{\z_f} \quad,\quad \z_B \equiv  \z_0+{\bar{\z}}
\eea
According to the general rule \rf{gr}:
\bea
\z'=\z_B+{\z_f}+\xi^\m \pa_\m (\z_B+{\z_f}) \la{zt1}
\eea
The transformed field $\z'$ can also be written
\bea
\z'\equiv \z_B+{\z_f}'  \la{zt2}
\eea
From \rf{zt1} and \rf{zt2} it follows
\bea
{\z_f}'={\z_f}+\xi^\m \pa_\m (\z_B+{\z_f})
\eea
Therefore the parameter $\xi$ satisfying $\xi^\m \pa_\m \z_B+{\z_f}+\xi^\m \pa_\m {\z_f}=0$ will lead to the gauge-fixing\footnote{More carefully one should consider
\bea
\z\equiv \z_B(x)+{\z_f}(y)
\eea 
and gauge away ${\z_f}(y)$. The argument $y$ of ${\z_f}(y)$ has been explicitly recorded to emphasize the 3D nature of the fluctuation ${\z_f}$.}
\bea
{\z_f}'=0
\eea
With ${\z_f}$ gauged away, the Feynman rules should be adopted accordingly: the Feynman diagrams with internal scalar loops become irrelevant. The discussion so far was for an arbitrary coordinate-dependent background. For a flat case we use the ``analytic continuation," namely, set $\z_0=const$, at the end \cite{Park:2015ybl}.

\subsubsection{scalar involving one-loop analysis}

Let us examine how the non-minimal coupling in Fig. 2 arises. As stated in section 2 there exist two diagrams in the conventional setup (i.e., the setup with the dynamical scalar) that produce the non-minimal coupling. The first diagram considered in the previous section becomes irrelevant with the gauge-fixing of the scalar. However, the second diagram, Fig. 2, remains relevant.
\bea
&&\begin{tikzpicture}[line width=1 pt, scale=1]
\draw (-140:1)--(0,0);
\draw (140:1)--(0,0);
\draw[vector] (.52,0) circle (.49cm);	

\node at (-140:1.2) {$\z_B$};
\node at (140:1.2) {$\z_B$};

\begin{scope}[shift={(1,0)}]
\draw[vector] (0:1)--(0,0);
\node at (0:1.3) {$\gt_B$};
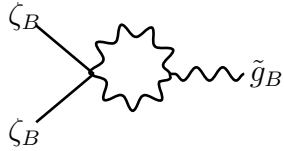
\node at (70:-1.4) {Figure 2: non-minimal coupling (graviton loop)};
\end{scope}
\end{tikzpicture} \nn
\eea
For the scalar-graviton four-point vertex the relevant terms come from the potential, i.e., the term quadratic in the scalar; the two graviton legs come from expanding the factor $\sqrt{-g}$. 
For the graviton three-point vertex the relevant terms are those in $\cL_{V_{I}},\cL_{V_{II}},$ and $\cL_{V_{III}},$ given in \rf{lv12} and \rf{gver}.
As stated in section 2 the divergence part can be computed by using a flat metric. The counterterm for the diagram in Fig. 2 can be obtained by evaluating
\bea
&& \fr{i^2}{2} \Big<2 \int(\cL_{V_{I}}+\cL_{V_{II}}+\cL_{V_{III}}) \;\int\Big(-\fr{{ \m_r^2}}{2}\Big) \sqrt{-g}\; \z_B^2 \Big> \nn\\
&\rightarrow& \fr{{ \m_r^2}}{16}  \z_B^2\Big< (\cL_{V_{I}}+\cL_{V_{II}}+\cL_{V_{III}})  (h^2-2h_{\m\n} h^{\m\n}) \Big>
\eea
The various numerical factors are the combinatoric factors.
The result turns out to be
\bea
\D \cL=\k'^2 \fr{{ \m_r^2}}{16} \fr{\G(\ve)}{(4\pi)^2}(-9\Rt)\z_B^2=- \k'^2 \fr{\G(\ve)}{(4\pi)^2} \fr{9 { \m_r^2}}{16} \;\z_B^2\,\Rt
\eea
In addition to the vacuum bubble diagram discussed in section 3.1 the diagram in Fig. 3 also contributes to renormalization of the cosmological constant. (In section 4 we will also see that this is one of the diagrams that make the cosmological constant time-dependent.)
\bea
\hspace{1.0in}
&&\begin{tikzpicture}[line width=1 pt, scale=1]
\draw (-140:1)--(0,0);
\draw (140:1)--(0,0);
\draw[vector] (.52,0) circle (.47cm);	

\node at (-140:1.4) {$\pa\z_B$};
\node at (140:1.4) {$\pa\z_B$};

\begin{scope}[shift={(1,0)}]
\draw (-40:1)--(0,0);
\draw (40:1)--(0,0);
\node at (-40:1.4) {$\pa\z_B$};
\node at (40:1.4) {$\pa\z_B$};	
\end{scope}
\end{tikzpicture}
\nn\\
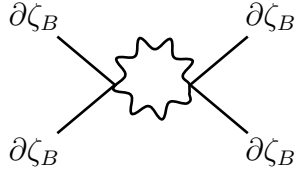
&&\hspace{-1.3in}\mbox{ Figure 3:  one-loop diagram with external scalar fields	} \nn
\la{4ptsc}
\eea
The correlator to compute is
\bea
&&\fr{i^2}2\Big<\Big[ -\fr12\int\Big(h^{\m\r}h_{\r}^\n+\fr18 \eta^{\m\n}(h^2-2h_{\r\s}h^{\r\s})-\fr12 hh^{\m\n}\Big) \pa_\m\z_B  \pa_\n \z_B \Big]^2   \Big> \nn\\
&=&-\fr18 \pa_\m\z_B  \pa_\n \z_B \;   \pa_{\m'}\z_B  \pa_{\n'} \z_B 
      \Big<\int\int\Big[h^{\m\r}h_{\r}^\n+\fr18 \eta^{\m\n}(h^2-2h_{\r\s}h^{\r\s})-\fr12 hh^{\m\n}\Big]  \nn\\
 &&  \hspace{.5in}   \Big[h^{\m'\r'}h_{\r'}^{\n'}+\fr18 \eta^{\m'\n'}(h^2-2h_{\r'\s'}h^{\r'\s'})-\fr12 hh^{\m'\n'}\Big]  \Big>
\eea
After some algebra one can show that
\bea
\D \cL=\fr{3}{16}(\k')^4\fr{\G(\ve)}{(4\pi)^2}\;(\pa^\m \z_B \,\pa_\m \z_B)^2  \la{s4pt}
\eea
The following diagram yields a vanishing result and thus need not be further considered\footnote{There are diagrams that belong solely to the renormalization of the scalar sector parameters. For example, the scalar coupling constant gets renormalized by the following loop effect:
	\bea
	&&\begin{tikzpicture}[line width=1 pt, scale=1]
	\draw (-140:1)--(0,0);
	\draw (-193:1)--(0,0);
	\draw (193:1)--(0,0);
	\draw (140:1)--(0,0);
	\draw[vector] (.52,0) circle (.49cm);	
	
	\node at (-140:1.3) {$\z_B$};
	\node at (-192:1.4) {$\z_B$};
	\node at (194:1.4) {$\z_B$};
	\node at (140:1.35) {$\z_B$};
	
	\begin{scope}[shift={(1,0)}]
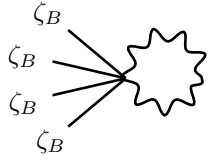
	\node at (70:-1.4) {Figure 5: scalar coupling renormalization};
	\end{scope}
	\end{tikzpicture} \nn
	\eea
We will not work out such diagrams here.	
	}:
\bea
&&\begin{tikzpicture}[line width=1 pt, scale=1]
\draw (-140:1)--(0,0);
\draw (140:1)--(0,0);
\draw[vector] (.52,0) circle (.49cm);	

\node at (-140:1.4) {$\pa\z_B$};
\node at (140:1.4) {$\pa\z_B$};

\begin{scope}[shift={(1,0)}]
\draw[vector] (0:1)--(0,0);
\node at (0:1.4) {$\gt_B$};
\node at (70:-1.4) {Figure 4: };
\end{scope}
\end{tikzpicture} \nn
\eea
We are now ready to pursue the main goal of the present work: the explicit implementation of one-loop renormalization of the gravity-scalar system.

\subsection{one-loop renormalization}

The one-loop renormaization of pure gravity without a cosmological constant was established long ago \cite{'tHooft:1974bx}. In the case of pure gravity it was crucial to use the following well-known topological identity: 
\bea
R_{\m\n\r\s}R^{\m\n\r\s}-4R_{\m\n}R^{\m\n}+R^2=\mbox{total derivative} \la{Riemannsqid}
\eea
When matter is present some of the counterterms cannot be absorbed even after using this identity (and field redefinitions).\footnote{This was the case for the pure gravity without the cosmological constant. Once the cosmological constant is included, the system becomes renormalizable with an appropriate metric field redefinition (originally due to 't Hooft \cite{'tHooft:1973us}) discussed below. The implication of such a field redeinition for the boundary condition has been recently investigated in \cite{Park:2016fxc}.}
Below we will collect all the counterterms obtained in the previous subsections and absorb them into a redefined metric. The gauging-away of the scalar fluctuation - which we view as a novel integrating-out procedure - makes the whole divergence removal procedure similar to that of pure gravity. It is only the cosmological constant-type terms that require extra care.

The scalar gauge-fixing has the following implications at a technical level that remain valid in the time-dependent background considered in the next subsection.
As we saw in section 3.2, the diagrams with internal scalar lines become irrelevant. Further, the gauge-fixing implies that what would be viewed as different counter terms due to different structures in the scalar field factors can now be grouped as one counterterm. The condition that the scalar is a constant can be used toward the final stage of the analysis after the effective action and the quantum-corrected field equations including that of the scalar are obtained. (The scalar field equation should be kept as a constraint.\footnote{The gauge-fixing of the fluctuating part of the scalar field means non-dynamism of the scalar {\em in the context of obtaining the 1PI action.} Once the 1PI action is obtained, one will have to go through the standard procedure (see e.g. \cite{Salopek:1988qh}) of gauge-fixing appropriate for the purpose at hand. At that point, the scalar should be taken dynamical just as it is in the classical level analysis.})

Let us collect all of the one-loop counterterms computed so far. (The analyses here and in section 4 are valid up to renormalization conditions. The gravity sector result including the ghost contribution was reviewed in section 2:
\bea
\D \cL 
&=& \fr12\fr{\G(\e)}{(4\pi)^2}\Big[\fr{41}{60}R_{\m\n}R^{\m\n}{ +{ \fr{1}{120}}}R^2\Big]\la{totalctr}
\eea
Here and below the tildes and subscript `$B$'s are omitted. A word of caution is in order. Although we are using the 4D covariant notation, the reduction to 3D is to be understood for the external states. Therefore, \rf{totalctr}, for example, can be explicitly further reduced in terms of $R_{mn}R^{mn}, R_{m3}R^{m3}$ and $R_{33}R^{33}$ that can be expressed in the 3D quantities (see, e.g., \cite{Aliev:2004ds} or \cite{Park:2015xoa}).\footnote{Strictly speaking, the renormalizability at one-loop is established with the help of the identify \rf{Riemannsqid} and it is not necessary to invoke the reduction scheme. It will be necessary do so in higher order loops.}

The counterterm associated with Fig. 2 takes the form of the non-minimal coupling and was shown to be
\bea
\D \cL=- \k'^2\fr{\G(\ve)}{(4\pi)^2} \fr{9\m^2}{16} \;\z^2\,R \la{nmq}
\eea
As for the renormalization of the cosmological constant there are two contributions: the bubble diagram contribution discussed in sec 3.1 and the contribution from Fig. 3.
The counterterm for the scalar 4-point amplitude was obtained in \rf{s4pt}:
\bea
\D \cL=\fr{3}{16}\k'^4\fr{\G(\ve)}{(4\pi)^2}\;(\pa^\m \z \,\pa_\m \z)^2  \la{s4ptq}
\eea
The counterterm \rf{nmq} implies that the Newton's constant becomes renormalized. (Once the scalar field is put onshell, it is just a constant, $\z=\sqrt{-\fr{\m^2}{\l}}$.) The couner-term \rf{s4ptq} vanishes once the scalar field is put onshell. 
Except for the generation of the corrections in the cosmological and Newton's constants, the one-loop renormalization of the system with a constant scalar background is not much different from that of the pure gravity system. Let us explicitly see how to absorb the divergences. 
Consider the Einstein-Hilbert with the following metric redefinition (the subscript $r$'s are omitted) {\cite{'tHooft:1973us}}, 
\bea
g_{\m\n}\ra g_{\m\n}+c_1 \k^2 g_{\m\n}R+c_2 \k^2R_{\m\n};  \la{ms}
\eea
the action becomes
\bea
\fr1{\k^2}\int d^4 x \sqrt{-g}\;R &\ra&  \fr1{(\k+\d \k)^2}\int d^4 x \sqrt{-g}\;R \nn\\
&&\hspace{-.2in}+\fr1{\k^2} \int d^4 x \sqrt{-g}\Big[(c_1+\fr12 c_2)R^2-c_2 R_{\m\n}R^{\m\n}\Big]
\eea
We should also consider the effect of \rf{ms} on cosmological constant-type term of the classical potential:
\bea
&&-(\fr{2}{\k^2}\L+V)\int \sqrt{-g}  \\
& \ra & -\Big[\fr{2}{(\k+\d \k)^2}(\L+\d \L)+V+\d V\Big] \int \sqrt{-g}\;\Big[1+\fr12 s_0\Big]\nn\\
&\ra & -\Big[\fr{2}{\k^2}\Big(\d\L-\fr{2\d\k \L}{{ \k}}\Big)+\d V\Big] \int \sqrt{-g} \;-\;\Big[\fr{2}{\k^2}\L+V\Big] \int \sqrt{-g}\;\Big(\fr12 s_0 \Big)\nn
\eea
 where
\bea
s_0 \equiv   \k^2(4c_1+c_2)R
\eea
$\d V$ has been introduced to match with the counter terms associated with $\l$-renormalization (Fig. 5) and $\m^2$-renormalization (a diagram similar to Fig. 5 but with two $\z_B$ legs). These diagrams will not be explicitly worked out.
Now the counterterms for the vacuum bubble discussed in section 3.1 and non-minimal term can be absrobed by setting 
\bea
\fr{2}{\k^2}\Big(\d\L-\fr{2\d\k  \L}{{ \k}}\Big)+\d V =  \mathscr{I}_{div}   \la{dL}
\eea
and 
\bea
-\fr{2}{{ \k^3}}\d\k   \;-\;\Big(\fr{2}{\k^2}\L+V\Big) \fr12 \k^2(4c_1+c_2)
= \fr{\G(\ve)}{(4\pi)^2}\Big(3m^2  - \k'^2 \fr{9\m^2}{16} \;\z^2 \Big)
\la{dk}
\eea
respectively. $\mathscr{I}_{div}$ is the parameter introduced in section 3.1. Eq. \rf{dk} determines $\d \k$; $\d \L$ is determined once that result is substituted into \rf{dL}. The counterterms in \rf{totalctr} can be absorbed by choosing $c_1,c_2$ as follows:
\bea
 c_1+\fr12 c_2={ \fr{1}{240}}\fr{\G(\e)}{(4\pi)^2} 
	\quad,\quad -c_2= \fr{41}{120}\fr{\G(\e)}{(4\pi)^2} \nn\\
\eea
which yields
\bea
c_1={ \fr{7}{40}}\fr{\G(\e)}{(4\pi)^2} \quad,\quad  c_2= -\fr{41}{120}\fr{\G(\e)}{(4\pi)^2} 
\eea	
With these $\d\k$ is given by 
\bea
\d \k =-\fr{\G(\e)}{(4\pi)^2}\k^3\Big[\fr{{ 43}}{240}\L+\cO(\k^2)\Big]
\eea
where we have substituted $m^2=\fr{\k^2\m^4}{4\l}, \z^2=-\fr{\m^2}{\l}$.
Similarly, one gets
\bea
 \d \L=\fr{\k^2}{2}(\mathscr{I}_{div}+\d V)-\fr{\G(\e)}{(4\pi)^2}\k^2\L\Big(\fr{{ 43}}{120}\L+\cO(\k^2)\Big)
 \eea
The procedure in this subsection will be generalized to the time-dependent background.

\section{Time-dependent background}

Let us pause and summarize what has been done. We have started with \rf{grv-sclrq3} and made the shift
\bea
g_{r\m\n}\equiv  h_{\m\n}+\tilde{g}_{{}_B\m\n}\quad \mbox{where}\quad \tilde{g}_{{}_B\m\n}\equiv \vf_{{}_B\m\n}+g_{0\m\n}   \la{gshiftq}
\eea
where the subscript $r$ on $g_{r\m\n}$ indicates that it is a renormalized field.
For the divergence analysis, we took $g_{0\m\n}=\eta_{\m\n}$ even if a curved solution is being considered. 
The renormalizability has been established by considering a constant scalar background. Things become technically more complicated once a time-dependent background and/or the finite parts are considered as we now turn. We do not quantitatively pursue the task of working out the finite parts. Instead, the renormalization procedure will be outlined by focusing on how to absorb the counterterms into the paramter/field redefinitions. After this task is completed the cosmological implications will be discussed.

\subsection{outline of renormailzation procedure}

Considering a constant background is sufficient for establishing renormalizability itself. However, one must consider the actual background of interest to study various physics associated with that background.
In the de Sitter case the reduction was carried out in the static coordinates \cite{Park:2015ybl}.\footnote{In the de Sitter case the 3D formalism of \cite{Park:2014noa} was used in \cite{Park:2015ybl} because of the infrared divergence issue. } For a more general Friedmann-Robertson-Walker (FRW) spacetime one may reduce along the genuine time coordinate $t$ \cite{Park:2014tia}.
It will be possible, although technically involved (mainly because of the complicated propagator of the background of interest), to obtain the 1PI action in the first few loop orders. The renormalization procedure of the previous subsection will go through only with relatively minor modifications. Renormalizability should be guaranteed by the constant background analysis although the details will go differently. One can renormalize the action, introduce bare quantities and obtain the 1PI action after path-integrating over $h_{\m\n}$. The path-integral over the scalar field will be unnecessary since its fluctuation part can be gauged away. Here again, the scalar gauge-fixing implies that what would be viewed as different counterterms due to the different scalar structures can be grouped as a single counterterm. One of the implications of this is that the diagram in Fig. 3 will contribute to the time-dependence of the vacuum energy.

The 1PI action will be a functional of $\tilde{g}_{{}_B\m\n}$ and ${\z}_B$. One may take the functional derivatives of the 1PI action with respect to $\tilde{g}_{{}_B\m\n}$ and ${\z}_B$ in order to obtain the loop-corrected field equations. (All the divergences will be cancelled automatically since the 1PI action contains the counterterms.) At this point the scalar field ${\z}_B$ can be set to a time-dependent function $\z_{0q}(t)$: ${\z}_B=\z_{0q}(t)$, and with this the loop-corrected scalar field equation can be viewed as the loop-corrected constraint. (The subscript $q$ stands for ``quantum.") The meaning of $\z_{0q}(t)$ is as follows.  Suppose the loop-corrected metric and scalar field equations are solved. $\z_{0q}(t)$ denotes the scalar solution.

The metric redefinition in \rf{ms} can be generalized to (again the index $r$ has been suppressed), 
\bea
g_{\m\n}\ra  \Big[g_{\m\n}+\k^2(c_0g_{\m\n}+c_1  g_{\m\n}R+c_2 R_{\m\n}+c_3 \pa_\m \z \pa_\n \z+c_4 g_{\m\n}(\pa\z)^2    )\Big] ;  \la{msg}
\eea
With this shift and the shift in the Newton's constant the Einstein-Hilbert part becomes
\bea
&&\hspace{.2in}\fr1{\k^2}\int d^4 x \sqrt{-g}\;R    \ra  \fr1{(\k+\d \k)^2}\int d^4 x \sqrt{-g}\;R \\
&&\hspace{-.6in}+ \fr1{(\k+\d \k)^2} \int d^4 x \sqrt{-g}\; \k^2\Big[c_0R+\Big(c_1+\fr12 c_2\Big)R^2-c_2 R_{\m\n}R^{\m\n}
 + \Big(\fr{c_3}{2}  - c_4\Big) R (\pa\z)^2 -c_3 R^{\m\n}\pa_\m\z \pa_\n \z \Big] \nn
\eea
Including the shift in the cosmological constant term one gets\footnote{We are considering things in the leading order of $\k$. At the subleading order one should consider the field redefinition of the metric contained in the matter part.}
\bea
&& \hspace{.8in}\fr1{\k^2}\int d^4 x \sqrt{-g}\;(R-2\L)    \\
&\ra&\hspace{.2in} \int d^4 x \sqrt{-g}\;\Big[\fr1{\k^2}(R-2\L) -2\Big(\fr{c_3}{2}+2c_4\Big)\L  \;(\pa\z)^2    \nn\\
&& \hspace{-.8in}-2\Big(\fr{\d\L}{\k^2}+2c_0\L -\fr{2\d\k }{\k^3}\L\Big) - \Big(\fr{2\d\k}{\k^3} { -c_0} +(4c_1+c_2)\L\Big)\;{ R}+\Big(1-\fr{2\d\k}{\k}\Big)(c_1+\fr12 c_2) \;R^2\nn\\
&& \hspace{-.7in}-\Big(1-\fr{2\d\k}{\k}\Big) c_2 \;R_{\m\n}^2+\Big(1-\fr{2\d\k}{\k}\Big) \Big(\fr{c_3}{2}  {+} c_4\Big) \;R(\pa\z)^2
-\Big(1-\fr{2\d\k}{\k}\Big) c_3 \;R^{\m\n}\pa_\m\z \pa_\n\z\Big]   \nn
\eea
The first term on the right-hand side is the classical renormalized action. The second term should be absorbed by the field renormalization of the scalar field and will not be considered further by the same token by which the diagram in Fig. 5 was not considered. The rest of the terms can be matched with the counterterms computed in the previous section, and this way the parameters $\d\k,\;\d\L$ and $c$'s can be determined. 
The counterterm of the cosmological constant form can be absorbed by\footnote{If the cosmological constant $\L$ was absent, the divergence of the cosmological constant form can be viewed as the corresponding bare term.}
\bea
2\Big( \fr{\d\L}{\k^2} -\fr{2\d\k  \L}{\k^3}+2c_0\L  \Big) =  \mathscr{I}   \la{dLtd}
\eea
The contribution in Fig. 3 is higher order in $\k'$ and has not been included here; once included it is one of the terms leading to the time-dependence of the vacuum energy. The shift in the Newton's constant can be determined by requiring
\bea
\fr{2}{\k}\d\k_r +\k^2(4c_1+c_2)\L= {\k'^2}\fr{\G(\ve)}{(4\pi)^2} \fr{9\m^2}{16} \;\z_B^2 \la{ntncon}
\eea
The counterterms in \rf{totalctr} can be absorbed by choosing $c$'s as follows:
\bea
&&	 c_1+\fr12 c_2={ \fr{1}{240}}\fr{\G(\e)}{(4\pi)^2} 
\quad,\quad -c_2= \fr{41}{120}\fr{\G(\e)}{(4\pi)^2} 
\eea
The vanishing result associated with Fig. 4 implies
\bea
&&  (1-\fr{2\d\k}{\k}) (\fr{c_3}{2}  - c_4)=	0\quad,\quad  -(1-\fr{2\d\k}{\k}) c_3=0
\eea
These relationships yield
\bea
c_1={ \fr{7}{40}}\fr{\G(\e)}{(4\pi)^2} \quad,\quad  c_2= -\fr{41}{120}\fr{\G(\e)}{(4\pi)^2} 
 \quad,\quad c_3=0  \quad,\quad c_4=0
\eea	
Upon substituting these into \rf{ntncon} and \rf{dLtd} one gets (the constant $c_0$ remains unfixed; it may be set to zero if it remains unnecessary in the higher loops)
\bea
\d\k_r &=&	\fr{\G(\e)}{(4\pi)^2}\fr{\k^3}{2}\Big[\fr98 \m^2\z^2(t)-\fr{{ 43}}{120}\L   \Big]\nn\\
\d \L &=& \k^2\Big(\fr{\mathscr{I}_{div}}2-2c_0 \L\Big)+\fr{\G(\e)}{(4\pi)^2}\ \k^2 \L\Big[\fr98 \m^2\z^2(t)-\fr{{ 43}}{120}\L\Big]
\eea

\subsection{cosmological implications}

Let us consider a time-dependent $vev$ of the scalar and study its implications for cosmology.  
Our scheme of quantization leads to several interesting outcomes. For example, we show that the vacuum energy should be time-dependent. There has been a proposal in the literature that the Higgs field may play a role of the inflaton field \cite{Bezrukov:2007ep,Barvinsky:2012zz,Hamada:2012bp}. Our results seem to indicate the possibility for the Higgs field to play the role of the quintessence field as well. At least, it is not inconsistent with the quintessence idea at the most basic level; the Higgs field may be one of the contributors to the time-dependent vacuum energy. The discussion here does not exclude the possibility of the existence of a separate Quintessence field. What remains true regardless is the fact that the Higgs field will contribute to the time dependence of the vacuum energy no matter how small the dependence may currently be.

Suppose one has obtained the 1PI action of the gravity-scalar system and the quantum-corrected field equations. Based on isotropy and homogeneity one would set the metric ansatz to
\bea
ds^2=-dt^2+a^2(t) d\vec{x}^2, 
\eea
namely, the same form of the ansatz used for the classical field equations. If the solution for the scalar has certain characteristics (to which we will shortly turn), its potential will behave as a time-varying cosmological constant, which is nothing but the basic idea of Quintessence.

Although the full quantum analyisis is desirable and could be carried out in spite of the high-level technicalities, it would be of limited experimental significance for the time being. As stated previously, however, it provides a rationale for the setup in which some of the unsolved cosmological problems can be tackled and more precise formulations of relatively recent new ideas such as the time-dependent fundamental constants can be made. With this said one can for now focus on the classical field equations.
With a slowly varying solution the scalar kinetic term can be neglected. Again the smallness of the cosmological constant may be explained by examining the potential once the scalar solution is substituted.

The time-dependence of the scalar field will be such that it will settle down, as time goes on, to the value that yields the minimum of the potential. The fact that the scalar field contributes to the time-varying vacuum energy can be easily seen by examining the metric field equation,
\bea
R_{r\m\n}-\fr12 g_{r\m\n}R=8\pi G\Big[\fr12 \pa_\m{\z}_r \pa_\n {\z}_r-\fr14 g_{r\m\n} (\pa \z_r)^2-\fr12 g_{r\m\n} V \Big]:
\eea
the scalar kinetic terms may be neglected if they are much smaller than the potential term that is time-varying once the time-dependent solution for the scalar field is substituted.

The differential equation satisfied by the scalar field can be solved numerically as we will now discuss. We have restored $c$ and $\hbar$ in the equations below for the numerical analysis.  
The Friedmann equation (for the zero curvature case) is 
\bea 
H^2=\fr{8\pi G}{3c^2} \Big(\fr12 \dot{\z}^2+V(\z) \Big)
\eea
where
\bea
V=\fr{\l}{4}\Big(\z^2+\fr{1}{\l} \m^2\Big)^2  
\eea
The scalar field equation reads
\bea
\ddot{\zeta}+3H \dot{\z}+V'(\z)=0
\eea
where
\bea
H=\sqrt{\fr{8\pi G}{3c^2}\Big(\fr12 \dot{\z}^2+V\Big)}
\eea
The Higgs particle in the Standard Model has the values of $\l \simeq 0.13$ and
\bea
m_{H}\simeq 126\; GeV, \quad v\simeq 246\; GeV
\eea
where $m_H$ denotes the mass of a Higgs particle. In the CGS units the order of magnitude of $v$ is
\bea
v=\sqrt{-\fr{\m^2}{\l}}\sim \fr{m_{H}c}{\hbar \sqrt{\l}} \sim  10^{16}\; erg
\eea
Because of limitations in our computing power we could not quite use the numbers of this order. Instead we set
\bea
v \sim 10^5
\eea
For the same reason we set 
\bea
\fr{8\pi G}{3c^2} \sim 10^{-7}
\eea
Numerical solutions with various intial conditions have been examined: they all display highly oscillating behaviors for early times.
For a sufficiently large time, however, the scalar field quickly approaches the value used for
\bea
\z=\sqrt{-\fr{\m^2}{\l}}
\eea

\section{Conclusion}

In this work we have carried out the one-loop renormalization of a gravity-scalar system with the Higgs-type potential. We have reviewed that the 4D covariance is maintained only with the traceless propagator but not with the traceful propagator; the pathology of the traceful propagator was demonstrated with several examples including the non-minimal coupling that played an important role in the Higgs inflation proposal. Gauging-away of the fluctuation of the scalar field brought simplification in the counter-vertex structures, which in turn has substantially simplified the renormalization procedure.  
Once a time-dependent solution was considered the time-dependence of the vacuum energy followed. The novel gauge-fixing of the scalar facilitates the view of the potential as the time-varying vacuum energy. (The gauge-fixing of the scalar provides a rationale for viewing the scalar field as the onshell background while keeping the metric offshell.) Our result is in line with the view that the current small value of the cosmological constant may be due to the old age of the Universe \cite{Peebles:2002gy,Weinberg4}.

The present renormalization program offers several insights. 
Intead of asking why the cosmological constant is so small one should perhaps ask why the minimum of the potential should be taken to be zero. It might not be entirely possible to ``explain" why the minimum should be zero but it may be possible to atribute it to an initial or boundary condition which may be posed as a postulate. One possible motivation for the the zero-minimum potential is that only in such a case would a flat spacetime remian as a solution. (With the classical scalar solution $\z=0$ one has a false vacuum of dS whereas the true vacuum is a Minkowski spacetime, at least classically.) This will be the case even at the quantum level with an appropriate renormalization condition. The renormalization condition should be such that the solution approaches a flat spacetime; this will mean that all of the finite parts are taken zero by the renormalization condition. (See more comments below.)

Another insight is on the possible time-dependence of fundamental constants. The time-dependence of the vacuum energy (and other fundamental constants such as the masses of elementary particles) comes from the fact that the background (i.e., the FRW metric) is time-dependent. Presumably the smallness of the cosmological constant means that the current universe is such that the scalar value is very close to the true minimum.

\vspace{.3in}

There are several future directions, some of which must be completed to (dis)prove the statements and speculations above.

Firstly, one may tackle the finite parts of the correlators. To that end one must employ the full curved space propagator. The full propagator is the inverse of the Laplacian operator constructed out of $\tilde{g}_{{}_B\m\n}$. The metric $\tilde{g}_{{}_B\m\n}$  and its inverse should be expanded around $\tilde{g}_{{}_0\m\n}$ in the first order in $\vf_{{}_B\m\n}$ for the leading order computation of the effective action. {For a fuller analysis of renormalization, the renormalization conditions will have to be explicitly spelled out as well; the procedures introduced in \cite{Pius:2014iaa}\cite{Carvalho:2013wsa}\cite{Pelissetto:2015yha} will be useful.}

Secondly, one should consider the quantum corrections to the constraints. It will be crucial to check whether or not the reduction picture will remain valid with the quantum corrections. The best approach for this may be for one to first complete the renormalization procedure and cast the effective action into a simpler form in terms of the redefined fields. It is conveivable that many terms in the effective action may be absorbed into the field redefinition. Although the steps are expected to be involved, the complementary mathematical picture offers one positive sign: the condition \rf{constronn} should remain valid (by gauge-fixing if necessary) even after the quantum corrections are taken into account.

For pure gravity at one loop it was possible to stay within the 4D covariant setup \cite{'tHooft:1974bx} due to the topological identity given in \rf{Riemannsqid}. At two- and and higher- loops one must employ the ``hybrid" setup in which the external states are taken onshell 3D and the offshell momenta in the loop are 4D since the identity \rf{Riemannsqid} is not applicable \cite{Park:2015ota,Park:2015xoa}. 
As we saw in this work, essentially the same strategy can be employed for the gravity-matter system if the number of the matter components is small. It would be necessary, however, to use a different method for a more general gravity-matter system. (See \cite{Park:2016fxc} for the recent progress along this line.) But before we get to this, let us muse on one curious aspect of our foliation-based quantization scheme.

One of the greatly appreciated lessons of the holographic dualities in general is the importance of the boundary degrees of freedom and their dynamics. The ``dual" boundary degrees of freedom relevant for our case are directly visible as part of the bulk system \cite{Park:2015xoa,Park:2015ybl}. (They are directly visible in the sense that they do not require any transformation to become visible. In contrast the gauge degrees of freedom in AdS/CFT may become visible after a certain ``dualization process" \cite{Hatefi:2012bp}.) Because of this one faces potential tension with the Dirichlet boundary condition: there is a possibility that the Dirichlet boundary condition might not be adequate once quantum gravitational effects set in. (This issue has been recently analyzed \cite{Park:2016fxc}.) In black hole physics, for example, the information may escape from the black hole to the asymptotic region \cite{Park:2013rm} where it may get stored.
The Dirichlet boundary condition makes the boundary degrees of freedom non-dynamical since it constrains the fields to die out at the boundary. 
This does not pose a problem for a non-gravitational system since the system has genuine onshell bulk physics. Not all seems well in the gravitational physics, however, in light of our proposal on the manner in which the boundary degrees of freedom become relevant: the potential incompatibility between the Dirichlet boundary condition and the non-trivial boundary dynamics calls for investigation. In contrast, the Neumann boundary condition should be fully compatible with the quantum effects.
It also seems plausible that the Neumann boundary condition may shed some light in the black hole information paradox. (The present quantization scheme may be applied to study the Firewall \cite{Almheiri:2012rt}\cite{Braunstein:2014nwa,Braunstein:2009my}, the original motivation for our recent works.) Since the Dirichlet boundary condtion will wipe out the boundary dynamics the information might be lost if it (or at least part of it) is to be stored at the boundary. The Neumann boundary condtion should be safer in this respect.

Coming back to the renormalization of a more general gravity-matter system, we can somewhat articulate the possibility put forth in \cite{Park:2015ybl}. 
First of all, one should investigate the possibility that the presence of matter fields might not actually lead to non-renormalizability. In other words it could be that the renormalizability may be determined largely by the gravity sector since, with a theory in which the matter fields are minimally coupled with gravity, it is the graviton vertices that contain multiple derivatives, therefore tend to lead to more divergent results.\footnote{Also, as speculated in \cite{Park:2015ybl} the virual boundary terms could be of some help in absorbing some of the counter terms. This is in addtion to the flexibility, in a slightly more general renormalizability requirement, of having several terms not present in the original action as counterterms. The flexibility of virtual boundary terms includes the freedom analogous to the renormalization conditions. In other words, different finite virtual boundary terms will be analogous to different renormalization conditions.} More work will be required to thoroughly check this speculation. 
One encouraging sign is that renormalizability does not seem lost even if one keeps the scalar dynamical in the present system. (See \cite{Park:2016fxc} for recent confirmation of this anticipation.) One would introduce an additional term  
\bea
\sim g_{\m\n}\z^2
\eea
in the metric shift \rf{msg} in order to separately take care of the right-hand side of \rf{ntncon}: instead of \rf{ntncon}, one will now have two conditions, one for the Newton's constant and the other for absorbing the divergence of the form $R \z^2$.
It seems to be a reasonable possibility that a similar analysis will work for a system with many matter fields, such as the Standard Model.\footnote{ 
When the external states are taken as the 3D physical states the effective action can eventually be written as a 3D form. 
 Initially the effective action will be written as a position space 4D integral expression but then subsequently it should be possible to reduce it to the corresponding 3D form. The coefficients will reflect the 4D divergent integral.
}

\vspace{.2in}
These anticipations and speculations will be worth more careful study in the future.




\newpage
\appendix

\renewcommand{\theequation}{A.\arabic{equation}}
\setcounter{equation}{0}
\section{Conventions and identities}

The signature is mostly plus: 
\bea
\eta_{\m\n}=(-,+,+,+)
\eea
All the Greek indices are four-dimensional
\bea
\a,\b,\g,...,\m,\n,\r...=0,1,2,3
\eea
and all the Latin indices are three-dimensional
\bea
a,b,c,...,m,n,r...=0,1,2
\eea
Our definitions of the Riemann tensor, Ricci tensor and Ricci scalar are
\bea
&&R^\r{}_{\s\m\n}\equiv \pa_\m \G^\r_{\n\s}-\pa_\n \G^\r_{\m\s}
+\G^\r_{\m\l}\G^\l_{\n\s}-\G^\r_{\n\l}\G^\l_{\m\s}\nn\\
&&R_{\m\n}\equiv R^\k{}_{\m\k\n}\quad,\quad    R\equiv R^\n_\n             
\eea
The fluctuation metric $\f_{\m\n}$ is defined through
\bea
g_{\m\n}\equiv g_{0\m\n}+\f_{\m\n}
\eea
The indices of $\f_{\m\n}$ are raised and lowered by $g^{0\m\n},g_{0\m\n}$.
For the refined application of the background field method, an additional shift 
$\f_{\m\n}\equiv \f_{{}_B\m\n}+h_{\m\n}$ was made:
\bea
g_{\m\n}=g_{0\m\n}+\f_{{}_B\m\n}+h_{\m\n}
\eea
The following shorthand notations were used in some places:
\bea
\f\equiv g^{0\m\n}\f_{\m\n}\quad,\quad h\equiv g^{0\m\n}h_{\m\n}
\eea
The graviton propagator is given by
\bea 
<\f_{\m\n}(x_1)\f_{\r\s}(x_2)>&=& P_{\m\n\r\s}\, \D(x_1-x_2) 
\eea
where, for the traceless propagator,\footnote{
	The traceful propagator is
	\bea
	P_{\m\n\r\s}\equiv \fr12\Big(g_{0\m\r}g_{0\n\s}+g_{0\m\s}g_{0\n\r}
	- g_{0\m\n}g_{0\r\s}\Big)
	\eea
}
\bea
P_{\m\n\r\s}\equiv \fr12\Big(g_{0\m\r}g_{0\n\s}+g_{0\m\s}g_{0\n\r}
- \fr12g_{0\m\n}g_{0\r\s}\Big);
\eea
it satisfies 
\bea
P_{\m\n\k_1\k_2}P^{\k_1\k_2}{}_{\r\s}=P_{\m\n\r\s}
\eea
The ghost propagator is given by
\bea
<C_\m(x_1)\Cb_\n(x_2)>&=&g_{0\m\n}\,\D(x_1-x_2)
\eea
For a flat background,
\bea
\D(x_1-x_2)=\int \fr{d^4k}{(2\pi)^4}\fr{e^{ik\cdot (x_1-x_2)}}{i k^2}
\eea
For the scalar kinetic term the propagator is
\bea
<\z(x_1)\z(x_2)>=\int \fr{d^4k}{(2\pi)^4}\fr{e^{ik\cdot (x_1-x_2)}}{i k^2}
\eea
In the leading order in a flat background
\bea
\G_{\b\g}^\a=\fr12 (\pa_\b h_\g^\a +\pa_\g h_\b^\a-\pa^\a h_{\b\g})
\eea
The traceful case is
\bea
R_{\m\n} &\simeq& \fr12 \Big(\pa^\r\pa_\n \vf_{\m\r}+\pa^\r\pa_\m \vf_{\n\r}-\pa^2 \vf_{\m\n}-\pa_\m\pa_\n \vf \Big) \nn\\
R&\simeq & \pa_\m\pa_\n \vf^{\m\n}-\pa^2 \vf \nn\\
R^2 &\simeq& \pa_{\m}\pa_{\n}\vf^{\m\n}\,\pa_{\r}\pa_{\s}\vf^{\r\s}-2 \pa^2 \vf \pa_\a\pa_\b \vf^{\a\b}+\pa^2 \vf \pa^2 \vf
\\
R_{\a\b}R^{\a\b} &\simeq& \fr14\Big[\pa^2 \vf^{\m\n}\,\pa^2 \vf_{\m\n}-2\pa^2 \vf^{\a\k}\pa_\k \pa_\s \vf_\a^\s
+2(\pa_{\m}\pa_{\n}\vf^{\m\n})^2  +(\pa^2\vf)^2-2 \pa^2 \vf\,  \pa_{\m}\pa_{\n}\vf^{\m\n}
\Big]
\nn
\la{covctrf}
\eea
The traceless case is (one can simply remove all the terms containing $\vf$)
\bea
R_{\m\n} &\simeq& \fr12 \Big(\pa^\r\pa_\n \vf_{\m\r}+\pa^\r\pa_\m \vf_{\n\r}-\pa^2 \vf_{\m\n} \Big) \nn\\
R&\simeq & \pa_\m\pa_\n \vf^{\m\n}\nn\\
R^2 &\simeq& \pa_{\m}\pa_{\n}\vf^{\m\n}\,\pa_{\r}\pa_{\s}\vf^{\r\s}
\\
R_{\a\b}R^{\a\b} &\simeq& \fr14\Big[\pa^2 \vf^{\m\n}\,\pa^2 \vf_{\m\n}-2\pa^2 \vf^{\a\k}\pa_\k \pa_\s \vf_\a^\s
+2(\pa_{\m}\pa_{\n}\vf^{\m\n})^2
\Big]  \la{covctr}
\nn
\eea

\renewcommand{\theequation}{B.\arabic{equation}}
\setcounter{equation}{0}

\section{Pathology of traceful propagator}

In the main body it was shown that the counterterms turn out to be covariant once the traceless propagator is used.
In contrast, use of the traceful propagator yields the results that cannot be re-written in covariant forms. 
Instead of \rf{diag} and \rf{cross} one gets
\[
\hspace{-.2in}\D \cL=-\fr12 < (\cL_{V_I}+\cL_{V_{II}})^2 > 
=-\fr12 \fr{\G(\ve)}{(4\pi)^2}\Big[-\fr{7}{12}R^2+\fr{11}{6} R_{\a\b}R^{\a\b}
-\fr7{12}(2R^2 \pa^2\vf+(\pa^2\vf)^2)
\Big]
\]
\bea
\hspace{-.2in}\D \cL=-< (\cL_{V_I}+\cL_{V_{II}})\cL_{V_{III}} >
=  -\fr12\fr{\G(\ve)}{(4\pi)^2}\Big[ \Big(4 R^{\a\b}\pa^2 \vf_{\a\b}+\fr53 R\,\pa^\a\pa^\b \vf_{\a\b} -\fr83 R\,\pa^2 \vf  \Big) \Big] \nn\\
\eea
Neither of these can be reexpressed in a covaraint form.
For the non-minimal coupling diagram in Fig. 1 the counterterms are given by
\bea
\fr\l{24} \fr{\G(\ve)}{(4\pi)^2}\Big(\pa_\m\pa_\n \vf^{\m\n}+\fr12 \pa^2 \vf\Big)\z_B^2
\eea
This cannot be expressed in terms of covariant quantities.

\renewcommand{\theequation}{C.\arabic{equation}}
\setcounter{equation}{0}

\section{Coordinate-free analysis}

The analysis in section 2.2 crucially depends on the identity \rf{liecocom} derived in \cite{Kobayashi1} in the coordinate-free framework.
Whereas the component analysis requires less formalism and technicality, the coordinate-free analysis provides a complementary understanding of the subtle relationship between the 3D and 4D tensors. For example, the 3D and 4D covariant derivatives are related by a pullback operation and the use of the coordinate-free setup makes evident the generality of the techniques employed. In addition, the coordinate-free setup makes it easier to connect with the more mathematically-oriented literature.

The commutator of the component forms of the Lie and covariant derivatives used in section 2.2 is a special case of the more general relationship \cite{Kobayashi1}:
\bea
[\mathscr{L}_{\bfX}, \N_{\bfY} ]{\bf T}=\N_{[\bfX,\bfY]}{\bf T} \la{liecocom}
\eea
where $\bfX, \bfY$ are coordinate-free notations for vector fields and ${\bf T}$ a tensor field. For the present case, the vector fields $\bfX, \bfY$ take
\bea
\bfX=\fr{\pa}{\pa x^3}, \quad \bfY=\fr{\pa}{\pa x^\a} \quad 
\eea 
in the coordinate base, and
\bea
{\bf T}={\boldsymbol h},
\eea
where the boldfaced symbol ${\boldsymbol h}$ is the coordinate-free notation for the 3D metric.
Let us introduce the coordinate-free version of $e_a^\a $ (see Appendix A for its definition) and denote it by ${\boldsymbol e}$ \cite{Gourgoulhon}:
\bea
e_a^\a   \Leftrightarrow {\boldsymbol e}
\eea

The 3D covariant derivative ${\bf D}$ is related to the 4D covariant derivative $\boldsymbol\nabla$ by
\bea
{\bf D}={\boldsymbol e}^*\,{\boldsymbol\nabla}  \la{3d4dcov}
\eea
where ${}^*$ on ${\boldsymbol e}^*$ denotes a pullback operation and is the standard mathematical notation. It is also common in the mathematical literature to use the same symbol to denote a pullback quantity:
\bea
{\boldsymbol h}\equiv {\boldsymbol e}^*\,  {\boldsymbol h}  \la{spbm}
\eea
The 3D metric compatibility condition takes
\bea
{\bf D}  {\boldsymbol h}=0
\eea
Let us also denote the extension of the 3D second fundamental form ${\bf K}$ to 4D by the same letter ((3.59) of \cite{Gourgoulhon}):
\bea
{\bf K}\equiv {\boldsymbol e}^*\,  {\bf K}
\eea
The component form of the second fundamental form, $K_{ij}$, is related to the 3D metric $h_{ij}$ by
\bea
\Big(\fr{\pa}{\pa x^3}-\mathscr{L}_{\boldsymbol N}\Big)h_{ij}=2n K_{ij}
\eea
where $\boldsymbol N$ denotes the coordinate-free form of the shift vector, and
\bea
\mathscr{L}_{\boldsymbol N}h_{ij}=D_i N_j+D_j N_i
\eea
The component expression $K_{mn}K^{mn}$ should translate into the following coordinate-free expression:
\bea
( {\bf C}_{13} {\bf C}_{24})({\bf K}\otimes \#{\bf K})=({\bf K}\otimes \#{\bf K})\Big(\fr{\pa}{\pa x^\m},\fr{\pa}{\pa x^\n}; dx^\r,dx^\s \Big)
\la{Kscf}
\eea
where ${\bf C}$'s denote appropriate contractions. For example, ${\bf C}_{13}$ denotes the contraction of $\fr{\pa}{\pa x^\m}$ in the first slot and $dx^\r$ in the third slot. The subscripts such as $1,3$ are often suppressed.
 $\#{\bf K}$ is defined as (see e.g. \cite{Fecko})
\bea
\#{\bf K}\sim  {\bf C} ({\bf h^{-1}}\otimes  ({\bf C} ({\bf h^{-1}}\otimes {\bf K}) )
\eea 
where {\bf C} denotes the obvious contractions, i.e., the contractions dictated by the component expression, $K_{mn}K^{mn}$; the symbol $\sim$ means that the relationship is valid up to a possible numerical factor that depends on one's convention. Let us consider the coordinate-free expression of $K_{mn}K^{mn}$, \rf{Kscf}.
Partially integrating, the covariant derivative in $\#{\bf K}$ comes to act on the other ${\bf K}$; focusing on the factor $\bf D {\bf K}$ one gets
\bea
\Rightarrow  {\bf D}  \mathscr{L}_{\pa_3}  {\boldsymbol h}&=& {\boldsymbol e }^* \boldsymbol\nabla\; \mathscr{L}_{\pa_3}  {\boldsymbol h}
={\boldsymbol e }^* \mathscr{L}_{\pa_3}  \boldsymbol\nabla  {\boldsymbol h}
= \mathscr{L}_{\pa_3} {\boldsymbol e }^* \boldsymbol\nabla\;  {\boldsymbol h}
= \mathscr{L}_{\pa_3} {\bf D}  {\boldsymbol h} \nn\\
&=& 0
\eea

\renewcommand{\theequation}{D.\arabic{equation}}
\setcounter{equation}{0}

\newpage

\end{document}